\documentclass[a4paper,11pt]{article}
\usepackage{jheppub} % for details on the use of the package, please see the JINST-author-manual
\usepackage{lineno}
\newcommand{\rcite}[1]{ref.~\cite{#1}}
\newcommand{\rcites}[1]{refs.~\cite{#1}}

\usepackage{amsmath, amssymb, amsfonts, amsthm}
\usepackage{mathtools}
\usepackage[normalem]{ulem}

\usepackage[usenames,dvipsnames,table]{xcolor}
\definecolor{mygreen}{rgb}{0,0.4,0}
\definecolor{myblue}{rgb}{0,0.0,0.4}
\definecolor{refrcolor}{rgb}{0,0.4,0}
\definecolor{cgreen}{rgb}{0,0.7,0}
\definecolor{ecolor}{rgb}{.52,.03,.06}
\definecolor{bgcolor}{rgb}{.96,.95,.80}
\definecolor{bgcolordark}{rgb}{.80,.80,.67}
\definecolor{faint}{rgb}{.80,.80,.80}

\usepackage{tcolorbox}
\tcbuselibrary{breakable, skins}

\theoremstyle{plain}
\newtheorem{theorem}{Theorem}
\newtheorem*{theorem*}{Theorem}

\theoremstyle{definition}
\newtheorem{example}[theorem]{Example}

\newtcolorbox{myproofbox}[1][]{
  enhanced,
  breakable,
  borderline west={2pt}{0pt}{faint},
  notitle,
  before skip=10pt,
  after skip=10pt,
  colback=white, 
  colframe=white,
  frame hidden,
  boxrule=0pt, 
  boxsep=0pt,
  sharp corners,
  left=8pt, right=0pt, top=1pt, bottom=0pt,
  fontupper=\small,
  #1
}

\makeatletter
  {\popQED\endtrivlist\end{myproofbox}\@endpefalse%
  \if@noskipsec\leavevmode\fi\noindent\ignorespacesafterend}
\makeatother

\newtcolorbox{myexamplebox}[1][]{%
  enhanced,
  breakable,
  borderline west={2pt}{0pt}{faint},
  notitle,
  before skip=10pt,
  after skip=10pt,
  colback=white, 
  colframe=white,
  frame hidden,
  boxrule=0pt, 
  boxsep=0pt,
  sharp corners,
  left=8pt, right=0pt, top=1pt, bottom=0pt,
  #1
}

\makeatletter
\renewenvironment{example}[1][]{%
  \begin{myexamplebox}\refstepcounter{theorem}%
  \topsep6\p@\@plus6\p@\relax%
  \trivlist%
  \item[\hskip\labelsep\textbf{Example \theexample\@addpunct{.}}]%
  }%
  {\endtrivlist\end{myexamplebox}\@endpefalse%
  \if@noskipsec\leavevmode\fi\noindent\ignorespacesafterend}
\makeatother

\usepackage[english]{babel}
\usepackage{hyphenat}

\allowdisplaybreaks[1]

\makeatletter
\def\mr@ignsp#1 {\ifx\:#1\@empty\else #1\expandafter\mr@ignsp\fi}%
\newcommand{\multiref}[1]{\begingroup%\let\protect\string%
	\xdef\mr@no@sparg{\expandafter\mr@ignsp#1 \: }%
	\def\mr@comma{}%
	\@for\mr@refs:=\mr@no@sparg\do{\mr@comma\def\mr@comma{,}\ref{\mr@refs}}%
	\endgroup}
\renewcommand{\eqref}[1]{(\multiref{#1})}
\makeatother

\makeatletter
\newcommand{\namedref}[2]{\hyperref[#2]{#1~\ref*{#2}}}%
\newcommand{\namedreff}[2]{\hyperref[#2]{#1\,\ref*{#2}}}%
\newcommand{\secref}{\namedreff{Section}}
\newcommand{\subsecref}{\namedreff{Subsection}}
\newcommand{\appref}{\namedref{Appendix}}

\newcommand{\exref}{\namedref{Example}}
\makeatother

\numberwithin{equation}{section}

\newcommand{\eqn}[1]{eq.~\eqref{#1}}

\newcommand{\eqns}[2]{eqs.~\eqref{#1} and~\eqref{#2}}

\makeatletter
\providecommand*{\shuffle}{%
	\mathbin{\mathpalette\shuffle@{}}%
}
\newcommand*{\shuffle@}[2]{%
	\sbox0{$#1\vcenter{}$}%
	\kern .15\ht0 % side bearing
	\rlap{\vrule height .25\ht0 depth 0pt width 2.5\ht0}%
	\raise.1\ht0\hbox to 2.5\ht0{%
		\vrule height 1.75\ht0 depth -.1\ht0 width .17\ht0 %
		\hfill
		\vrule height 1.75\ht0 depth -.1\ht0 width .17\ht0 %
		\hfill
		\vrule height 1.75\ht0 depth -.1\ht0 width .17\ht0 %
	}%
	\kern .15\ht0 % side bearing
}
\makeatother

\newif\ifnote 
\notetrue

\title{\boldmath
Associators for AdS string amplitude building blocks
}

\usepackage{mathdots}

\newcommand{\Sb}{\overline{S}}
\newcommand{\zb}{\overline{z}}

\newcommand{\cL}{\mathcal{L}}

\newcommand{\cC}{\mathcal{C}}
\newcommand{\cO}{\mathcal{O}}
\newcommand{\cF}{\mathcal{F}}
\newcommand{\cJ}{\mathcal{J}}
\newcommand{\cI}{\mathcal{I}}

\newcommand{\Li}{\operatorname{Li}}
\renewcommand{\Re}{\operatorname{Re}}

\newcommand{\sv}{\operatorname{sv}}
\newcommand{\zetasv}{\zeta^{\sv}}
\newcommand{\cZ}{\mathcal{Z}}
\newcommand{\cZsv}{\mathcal{Z}^{\sv}}

\author{Konstantin Baune}
\affiliation{Institute for Theoretical Physics, ETH Zurich\\
Wolfgang-Pauli-Str.~27, 8093 Zurich, Switzerland}

\emailAdd{baunek@ethz.ch}

\abstract{We show that building blocks for open- and closed-string amplitudes on AdS are generated by the Drinfeld and Deligne associator, respectively. Our formalism lifts the known associator recursions for flat-space string amplitudes to the AdS picture. This delivers another proof that the AdS building blocks admit low-energy expansions with (single-valued) multiple zeta values as coefficients and provides all-order relations for the integral expressions.}

\begin{document}
\maketitle
\flushbottom

%%%%%%%%%%%%%%%%%%%%%%%%%%%%%%%%%%%%%%%
\section{Introduction}
\label{sec:intro}
Scattering amplitudes in quantum field theory and string theory are the simplest observables, whose analytic and algebraic properties need to be understood. When considering possible four-point amplitudes with infinite towers of mediated particles, Veneziano wrote down the \emph{Veneziano amplitude} as the Euler beta function~\cite{Veneziano:1968yb}
\begin{equation}\label{eqn:Veneziano}
	\beta(s,t)=\int_{0}^1dx\,x^{s-1}(1-x)^{t-1}=\frac{\Gamma(s)\Gamma(t)}{\Gamma(s+t)},
\end{equation}
which led to the birth of string theory and describes the four-point open-string tree-level amplitude. Later, also the four-point expression for the closed string was found, known as the \emph{Virasoro--Shapiro amplitude}, which can be written as the complex beta function~\cite{Virasoro:1969me,Shapiro:1970gy}
\begin{equation}\label{eqn:VS}
	\beta_{\mathbb{C}}(s,t)=-\frac{1}{2\pi i}\int_{\mathbb{C}}dz\,d\zb\,|z|^{2(s-1)}|1-z|^{2(t-1)}=\frac{\Gamma(s)\Gamma(t)\Gamma(1-s-t)}{\Gamma(s+t)\Gamma(1-s)\Gamma(1-t)}.
\end{equation}

Comparing the two integral expression for $\beta$ and $\beta_{\mathbb{C}}$, one might suspect a relation: The most well-known identity between them is the Kawai--Lewellen--Tye (KLT) relation~\cite{Kawai:1985xq}, which allows to express $n$-point closed-string tree-level amplitudes as ``squares'' of $n$-point open-string amplitudes. The four-point KLT relation reads~\cite{Kawai:1985xq,Brown:2019wna}
\begin{equation}
	\beta_{\mathbb{C}}(s,t)=-\frac{1}{2\pi i}\left(\frac{2\sin(\pi s)\sin(\pi t)}{i\,\sin(\pi(s+t))}\right)\beta(s,t)^2.
\end{equation}
This identity was later related to intersection theory \cite{Mizera:2017cqs,Mizera:2019gea} and also generalized for one-loop string amplitudes \cite{Stieberger:2022lss,Stieberger:2023nol,Mazloumi:2024wys,Mafra:2018qqe}.
Another way to formulate the relation between open- and closed-string amplitudes is through the \emph{single-valued map}. The simplest installation becomes visible when considering the low-energy expansions (i.e.~small Mandelstam variables $s,\,t$) of the above amplitudes (see e.g.~\rcites{Stieberger:2013wea,Brown:2019wna})
\begin{subequations}
\begin{align}
	\beta(s,t)&=\left(\frac{1}{s}+\frac{1}{t}\right)\exp\Bigg[-\sum_{n\geq2}\frac{\zeta(n)}{n}\left(s^n+t^n-(s+t)^n\right)\Bigg],\\
	\beta_{\mathbb{C}}(s,t)&=\left(\frac{1}{s}+\frac{1}{t}\right)\exp\Bigg[-\sum_{\substack{n\geq2\\n\text{ odd}}}\frac{2\zeta(n)}{n}\left(s^n+t^n-(s+t)^n\right)\Bigg],
\end{align}
\end{subequations}
where one notices that
\begin{equation}
	\zeta(2n)\mapsto0,\quad\zeta(2n{-}1)\mapsto2\zeta(2n{-}1),\quad n\in\mathbb{N},
\end{equation}
maps $\beta(s,t)$ to $\beta_{\mathbb{C}}(s,t)$. These are simple instances of the single-valued map for multiple zeta values, established in \rcites{BrownSVMPL,Brown:2013gia} and connected to tree-level string amplitudes of all multiplicities in \rcites{Stieberger:2013wea,Stieberger:2014hba}. On the level of the full integrals, this relation was proven by the means of single-valued integration \cite{Brown:2018omk,Brown:2019wna}, so that the integral $\beta_{\mathbb{C}}(s,t)$ is the single-valued image of $\beta(s,t)$ through a mapping of twisted (co)homologies.

In recent years, the connection of string amplitudes to polylogarithms and the Knizhnik--Zamolodchikov equation was studied thoroughly. This included finding that the Drinfeld associator, which can be represented as the generating series of multiple zeta values, recursively generates all tree-level open-string amplitudes~\cite{Broedel:2013aza,BK,Kaderli}. Correspondingly, the Deligne associator, as the generating series of all single-valued multiple zeta values, recursively generates the closed-string tree-level amplitudes~\cite{Stieberger:2013wea,Stieberger:2014hba,Brown:2019wna,Baune:2024uwj}.

While the structure of tree-level amplitudes in flat-space string theory is by now well understood, the interest in recent years was extended to considering string-amplitudes on AdS backgrounds. This was first established through a bootstrap program to find the large AdS radius corrections to the Virasoro--Shapiro amplitude on $AdS_5\times S^5$ \cite{Alday:2022uxp,Alday:2022xwz,Alday:2023jdk,Alday:2023mvu,Alday:2024xpq}. It was continued for other theories~\cite{Alday:2024rjs,Chester:2024esn,Chester:2024wnb,Wang:2025pjo} as well as for the AdS corrections to the Veneziano amplitude \cite{Alday:2024ksp,Alday:2024yax}.

One might wonder, how the KLT relations and the Drinfeld and Deligne recursions found in flat space generalize when considering the corresponding amplitudes on AdS. For this, one has to consider the types of integrals appearing for the curvature corrections, which are (complex) beta function integrals with insertions of (single-valued) polylogarithms in the integrand. In \rcite{Alday:2024yax} it was already shown, that in the open-string case, the low-energy expansion coefficients are zeta values. Correspondingly in \rcite{Alday:2023jdk}, it was proven that the coefficients are single-valued zeta values in the closed-string case. Very recently, \rcite{Alday:2025bjp} found that these integrals also obey double-copy relations, generalizing the KLT relation to the AdS integrals.\\
In \rcite{Alday:2025bjp} the question was raised, which types of associators correspond to AdS string amplitudes and their building blocks in the open- and closed-string cases. The current article aims to answer this question.

\paragraph{Summary.} We will show how the AdS amplitude building blocks for open and closed strings can be obtained through recursions using the Drinfeld associator and the Deligne associator, respectively, in suitable matrix representations. This yields all order expansions of the building blocks in the low-energy limit of small Mandelstam variables. Additionally, it provides another proof that the coefficients of these expansions are multiple zeta values for open strings and single-valued multiple zeta values for closed strings. Furthermore, our construction indicates that similar to flat space, Drinfeld and Deligne associators generate four-point open- and closed-string amplitudes on AdS.\\
Finally, we will present how these structures generalize for iterated integrals with polylogarithm insertions, reminiscent of higher-point amplitudes.

\paragraph{Outline.} The article is structured as follows. In \secref{sec:assrec} we review string integral recursions using the Drinfeld and Deligne associators, respectively. An introduction to four-point AdS amplitude building blocks for open and closed strings is provided in \secref{sec:adsamps}. In \secref{sec:assrel} we are going to present and study the new associator relations for the AdS building blocks of open and closed strings. Finally, we will discuss further research directions in \secref{sec:conclusions}.\\ 
Three appendices supplement this article: In \appref{app:polylogs} we summarize the basics and fix our conventions for multiple polylogarithms, multiple zeta values and their single-valued analogs, as well as the Drinfeld and Deligne associator. \appref{app:explicit3} collects results for the open-string Drinfeld associator relations of weight three AdS amplitude building blocks, while \appref{app:5pt} contains details for a higher-point example.

%%%%%%%%%%%%%%%%%%%%%%%%%%%%%%%%%%%%%%%

\section{Drinfeld and Deligne recursions}\label{sec:assrec}
The \emph{Drinfeld associator} $\Phi$ and the \emph{Deligne associator} $\Phi^{\sv}$---as reviewed in \appref{app:polylogs}---are the generating series of \emph{multiple zeta values} (MZVs) and \emph{single-valued multiple zeta values} (svMZVs), respectively,
\begin{subequations}\label{eqn:associators}
\begin{align}
	\Phi(e_0,e_1)&=\sum_{w\in\{e_0,e_1\}^\times}(-1)^{\#w}w\,\zeta_w,\\
	\Phi^{\sv}(e_0,e_1)&=\sum_{w\in\{e_0,e_1\}^\times}(-1)^{\# w}w\,\zeta^{\sv}_w=\sv(\Phi).
\end{align}
\end{subequations} 
They are valued in the free algebra generated by the letters $e_0,\,e_1$, where $\{e_0,e_1\}^\times$ refers to the set of all words in $e_0$ and $e_1$ including the empty word $e$ and we have the natural mapping $\zeta_{w}=\zeta_{\widetilde{w}}$ with $w\,{=}\,\prod_{j=1}^ke_0^{a_j}e_1^{b_j}\mapsto \widetilde{w}\,{=}\,\prod_{j=1}^{k}0^{a_j}1^{b_j}$ for (sv)MZVs. The expression $\# w$ in \eqn{eqn:associators} refers to the number of $e_1$'s appearing in a word $w$. Our conventions for polylogarithms and multiple zeta values agree with \rcite{Alday:2025bjp} and are collected in \appref{app:polylogs}. Let us furthermore define $\cZ=\mathbb{Q}[(\zeta_w)_{w\in\{0,1\}^\times}]$ and $\cZsv=\mathbb{Q}[(\zeta_w^{\sv})_{w\in\{0,1\}^\times}]\subset\cZ$ as the rings over $\mathbb{Q}$ generated by MZVs and svMZVs, respectively.

The above associators can be used to recursively generate and calculate integrals of Selberg type, such as string amplitudes, as follows. Let $F(x)$ be a meromorphic vector of iterated integrals, solving the \emph{Knizhnik--Zamolodchikov (KZ) equation}
\begin{equation}\label{eqn:kzf}
	\frac{d}{dx}F(x)=\left(\frac{e_0}{x}+\frac{e_1}{x-1}\right)F(x),
\end{equation}
for square matrices $e_0,\,e_1$. Then the regularized boundary values 
\begin{equation}\label{eqn:boundval}
C_0=\lim_{x\to0}x^{-e_0}F(x),\qquad C_1=\lim_{x\to1}(1-x)^{-e_1}F(x)
\end{equation}
are related by the Drinfeld associator via (see e.g.~\cite{Brownhyperlogs})
\begin{equation}\label{eqn:Drinrec}
	C_1=\Phi(e_0,e_1)\,C_0,
\end{equation}
which we will call a \emph{Drinfeld recursion} for a given representation of the triple $(F,e_0,e_1)$. The calculation of $C_0$ and $C_1$ usually comes alongside a limit prescription for certain parameters related to the extra point $x$.
This method was used for Selberg integrals in \rcite{Terasoma} and then applied to open-string amplitudes in \rcite{Broedel:2013aza} relating $(n{-}1)$-point amplitudes in $C_0$ to $n$-point amplitudes in $C_1$. The Drinfeld recursion for string amplitudes was further studied in the articles \cite{BK,Kaderli,Baune:2024uwj} and it provides a proof that the low-energy expansions of open-string integrals have coefficients valued in~$\cZ$.

If one however has a vector $\cF(z)$ of iterated integrals, single-valued in the extra variable $z$ and a solution to \eqn{eqn:kzf}, then the regularized boundary values $\cC_0\,{=}\,\lim_{z\to0}|z|^{-2e_0}\cF(z)$ and $\cC_1\,{=}\,\lim_{z\to1}|1{-}z|^{-2e_1}\cF(z)$ are related through the Deligne associator as \cite{Baune:2024uwj}
\begin{equation}\label{eqn:Delinrec}
	\cC_1=\Phi^{\sv}(e_0,e_1)\,\cC_0.
\end{equation}
For a given representation $(\cF,e_0,e_1)$, we will call \eqn{eqn:Delinrec} a \emph{Deligne recursion}, in analogy to the Drinfeld recursion \eqref{eqn:Drinrec}.
This technique was used in \rcite{Baune:2024uwj} to build a recursion for closed-string tree-level amplitudes, relating $(n{-}1)$-point amplitudes in $\cC_0$ to $n$-point amplitudes in $\cC_1$. The relation of closed-string amplitudes to the Deligne associator was already noticed in \rcite{Stieberger:2013wea}, which also implies that the low-energy expansions of closed-string integrals have coefficients valued in $\cZsv$.

In the next section we will review the building blocks for AdS string amplitudes before finding Drinfeld and Deligne recursions for the AdS building blocks in \secref{sec:assrel}.

%%%%%%%%%%%%%%%%%%%%%%%%%%%%%%%%%%%%%%%
\section{AdS string amplitude building blocks}
\label{sec:adsamps}
In this section we are going to review the structure of the integrals that constitute the building blocks of the AdS amplitudes for open and closed strings.

\subsection{Open-string AdS amplitudes}
For the study of four-point open-string AdS amplitudes at tree-level, the authors of \rcite{Alday:2025bjp} introduced the set of Euler beta integrals with a polylogarithm insertion,
\begin{equation}
\label{eqn:Jint}
	J_w^{s,t}=\int_0^1dy\,y^{s-1}(1-y)^{t-1}L_w(y),
\end{equation}
where the word $w\,{\in}\,\{0,1\}^\times$ labels the \emph{multiple polylogarithm}\footnote{See \appref{sec:MPLs} for definition and conventions of MPLs.} (MPL) $L_w$ in the integrand.
The above integrals are the basic building blocks of the amplitudes found in \rcites{Alday:2024yax,Alday:2024ksp}, and we will refer to them as $J$-integrals in the remainder of this article. In \rcite{Alday:2024yax} it was shown that these integrals admit expansions in small $s$ and $t$ with coefficients valued in $\cZ$. The goal of \secref{sec:assrel} will be to find associators that produce these $J$-integrals.

Let us look at three important properties of the $J$-integrals, derived in \rcite{Alday:2025bjp}.
First, using partial fractioning, one finds (c.f.~\cite[eq.~(2.12)]{Alday:2025bjp})
\begin{align}
\begin{split}
	J_w^{s,t}&=\int_0^1dy\,y^{s}(1-y)^{t}L_w(y)\frac{1}{y}\frac{1}{1-y}\\
	&=\int_0^1dy\,y^{s}(1-y)^{t}L_w(y)\left(\frac{1}{y}+\frac{1}{1-y}\right)\\
	&=J_w^{s,t+1}+J_w^{s+1,t}.
\end{split}
\end{align}
Second, since the space of MPLs $L_w(y)$ is closed under the transformation $y\,{\mapsto}\,1\,{-}\,y$, swapping $s$ and $t$ in $J_w^{s,t}$ yields back (a linear combination of) integrals $J_{w'}^{s,t}$, i.e.~the $J$-integrals are closed under the crossing operation $s\,{\leftrightarrow}\, t$.

Finally, an important property for shifts in the arguments $s$ and $t$ was found in \rcite{Alday:2025bjp}: If one writes the generating series of all $J$-integrals as
\begin{align}\label{eqn:Jgen}
\begin{split}
	\cJ(s,t;e_0,e_1)&=\sum_{w\in\{e_0,e_1\}^\times}w\,J_w^{s,t}\\
	&=J_e^{s,t}+e_0J_0^{s,t}+e_1J_1^{s,t}+e_0^2J_{00}^{s,t}+e_0e_1J_{01}^{s,t}+\ldots,
\end{split}
\end{align}
with non-commutative letters $e_0,\,e_1$, the shift relations\footnote{The multiplication by $s$ and $t$ in this equation is understood as the multiplication by $s\cdot e$ and $t\cdot e$ in the algebra spanned by $e_0$ and $e_1$, for the empty word $e$ acting as an identity element under concatenation.} \cite{Alday:2025bjp}
\begin{subequations}\label{eqn:shiftrel}
\begin{align}
	(s+t+e_0+e_1)\,\cJ(s+1,t;e_0,e_1)&=(s+e_0)\,\cJ(s,t;e_0,e_1),\\
	(s+t+e_0+e_1)\,\cJ(s,t+1;e_0,e_1)&=(t+e_1)\,\cJ(s,t;e_0,e_1)
\end{align}
\end{subequations}
hold, which will be of use later.

\subsection{Closed-string AdS amplitudes}

For the closed-string four-point amplitude on AdS backgrounds studied in the series of articles \cite{Alday:2022uxp,Alday:2022xwz,Alday:2023jdk,Alday:2023mvu,Chester:2024wnb,Chester:2024esn,Alday:2024rjs,Alday:2024xpq,Wang:2025pjo}, a general structure of the AdS curvature corrections to the Virasoro--Shapiro amplitude was identified as complex beta function integrals with insertions of \emph{single-valued polylogarithms}\footnote{Definition of svMPLs and the conventions used here are summarized in \appref{sec:svpolylogs}.} (svMPLs) $\cL_w$. This led the authors of \rcites{Alday:2023jdk,Alday:2025bjp} to define the closed-string building blocks
\begin{equation}\label{eqn:Iint}
	I_w^{s,t}=\int_{\mathbb{C}}d^2z\,|z|^{2(s-1)}|1-z|^{2(t-1)}\cL_w(z).
\end{equation}
These integrals are the single-valued images of the $J$-integrals, $I_w^{s,t}\,{=}\,\sv(J_w^{s,t})$ \cite{Alday:2025bjp}, and we will refer to them in the remainder of this article as $I$-integrals. In \rcite{Alday:2023jdk} these integrals have been evaluated as expansions in $s$ and $t$ using the tools of \rcite{VanZerb}, finding that the expansion coefficients are valued in $\cZ^{\sv}$.

Combining all $I$-integrals into the generating function
\begin{align}\label{eqn:Igen}
\begin{split}
	\cI(s,t;e_0,e_1)&=\sum_{w\in\{e_0,e_1\}^\times}w\,I_w^{s,t}\\
	&=I_e^{s,t}+e_0I_0^{s,t}+e_1I_1^{s,t}+e_0^2I_{00}^{s,t}+e_0e_1I_{01}^{s,t}+\ldots,
\end{split}
\end{align}
\rcite{Alday:2025bjp} shows the single-valued shift relation
\begin{subequations}\label{eqn:svshiftrel}
\begin{align}
	(s+t+e_0+e_1)\,\cI(s+1,t;e_0,e_1)\,(s+t+e_0+e_1')&=(s+e_0)\,\cI(s,t;e_0,e_1)\,(s+e_0),\\
	(s+t+e_0+e_1)\,\cI(s,t+1;e_0,e_1)\,(s+t+e_0+e_1')&=(t+e_1)\,\cI(s,t;e_0,e_1)\,(t+e_1'),
\end{align}
\end{subequations}
where $e_1'$ is a letter from a second alphabet, admitting an expansion in $e_0$ and $e_1$ as described in \appref{sec:svpolylogs}.

%%%%%%%%%%%%%%%%%%%%%%%%%%%%%%%%%%%%%%%
\section{Associator relations for AdS amplitude building blocks}
\label{sec:assrel}
In this section we will derive associator relations for the building blocks of AdS string amplitudes. This will be achieved by finding appropriate vectors $F(x)$ and matrix representations of $e_0$ and $e_1$ for a Drinfeld recursion as in \eqn{eqn:Drinrec} for the open-string AdS building blocks. Afterwards, we use the single-valued map to find corresponding Deligne recursions \eqref{eqn:Delinrec} for the closed-string AdS building blocks. Accordingly, the Drinfeld associator generates the open-string building blocks, while the Deligne associator produces the closed-string amplitude building blocks.

\subsection{Open-string building blocks and the Drinfeld associator}\label{sec:openrec}

As a starting point, we define a class of functions for the Drinfeld recursion of the open-string AdS amplitude building blocks. These functions are Selberg integrals (as used in the recursions in \rcites{Terasoma,Broedel:2013aza,BK,Kaderli,Baune:2024uwj}) with an extra insertion of a polylogarithm, generalizing the integrals $J_w^{s,t}$ from \eqn{eqn:Jint}. They are defined for a word $w\,{\in}\,\{0,1\}^\times$ and $x\,{\in}\,]0,1[$ by
\begin{equation}
\label{eqn:adsSelberg}
	S_w[i](x)=\int_0^xdy\, y^s(1-y)^t (x-y)^r \frac{L_w(y)}{y-y_i}.
\end{equation}
Note, that we omitted the dependence on the parameters $s,t$, which in the physical context are the Mandelstam variables describing the kinematics of the scattering process, and on the auxiliary parameter $r$. This parameter $r$ will finally be taken to zero for the calculation of the boundary values $C_0$ and $C_1$ of the Drinfeld recursion \eqref{eqn:Drinrec}. For the integral \eqref{eqn:adsSelberg} to converge, we take\footnote{In fact the region of convergence is larger depending on the choice of $i$, see discussions in \rcite{VanZerb} for the case without polylogarithm insertions.} $\Re r,\Re s,\Re t>0$. The index $i$ in \eqn{eqn:adsSelberg} can take values in $\{0,1,2\}$ corresponding to the points $y_0\,{=}\,0,\,y_1\,{=}\,1,\,y_2\,{=}\,x$.

The goal is to find a KZ equation for a vector of integrals of the type \eqref{eqn:adsSelberg}, so that the Drinfeld associator will relate the regularized boundary values for $x\,{\to}\,0$, $x\,{\to}\,1$, giving insights into the structure of the $J$-integrals. To do so, we first calculate the derivative of $S_w[i](x)$ w.r.t.~$x$, where we limit our attention to the cases $i\,{=}\,0,1$, since the case $i\,{=}\,2$ can be related to the others as we will see below.
\begin{align}
\label{eqn:derS}
\begin{split}
	\frac{d}{dx}S_w[i](x)&=\int_0^xdy\,y^s(1-y)^t(x-y)^r\frac{L_w(y)}{y-y_i}\frac{r}{x-y}\\
	&=\frac{r}{x-y_i}\int_0^xdy\,y^s(1-y)^t(x-y)^r L_w(y)\left(\frac{1}{y-y_i}-\frac{1}{y-x}\right)\\
	&=\frac{r}{x-y_i}(S_w[i](x)-S_w[2](x)).
\end{split}
\end{align}
In this calculation, we first notice that the contribution from the derivative of the integration boundary vanishes, since the integrand vanishes for $y\,{\to}\, x$, and second, we used partial fractioning ($i\,{\neq}\,2$)
\begin{equation}
	\frac{1}{y-y_i}\frac{1}{x-y}=\frac{1}{x-y_i}\left(\frac{1}{y-y_i}-\frac{1}{y-x}\right).
\end{equation}
Next, for a word $w\,{=}\,w_1\cdots w_n\in\{0,1\}^\times$ we will relate the three integrals $S_w[i](x)$, $i\,{=}\,0,1,2$, through the use of a total derivative:
\begin{align}
\label{eqn:Srel}
\begin{split}
	0&=\int_0^xdy\frac{d}{dy}y^s(1-y)^t (x-y)^r L_w(y)\\
	&=\int_0^xdy\,y^s(1-y)^t (x-y)^r L_w(y)\left[\frac{s}{y}+\frac{t}{y-1}+\frac{r}{y-x}\right]\\
	&\quad+\int_0^xdy\,y^s(1-y)^t (x-y)^r \frac{L_{w_2\cdots w_n}(y)}{y-y_{w_1}}\\
	&=s\, S_w[0](x)+t\, S_w[1](x)+r\, S_w[2](x)+S_{w_2\cdots w_n}[w_1](x)\\
	&=\Sb_w[0](x)+\Sb_w[1](x)+\Sb[2]_w(x)+S_{w_2\cdots w_n}[w_1](x),
\end{split}
\end{align}
where in the last line we defined $\Sb_w[i](x)\,{=}\,c_i\, S_w[i](x)$ for $c_0\,{=}\,s,c_1\,{=}\,t,c_2\,{=}\,r$.
In contrast to the derivatives found for the flat-space integrals considered in \rcites{Broedel:2013aza,BK,Kaderli,Baune:2024uwj}, here the total derivative also involves an integral with the insertion of a polylogarithm which is one letter shorter. Thus, in order to find a system of functions closed under taking derivatives, we will also have to consider integrals with shorter polylogarithm insertions, implying that a larger vector $F(x)$ is needed for the Drinfeld recursion \eqref{eqn:Drinrec}.

Combining \eqns{eqn:derS}{eqn:Srel} yields
\begin{equation}
		\frac{d}{dx}S_w[i](x)=\frac{1}{x-y_i}\left(r\, S_w[i](x)+\Sb_w[0](x)+\Sb_w[1](x)+S_{w_2\cdots w_n}[w_1](x)\right).
\end{equation}
Since $r$ is an auxiliary parameter and we want to recover integrals without the insertion $(x\,{-}\,y)^r$ in the end, we can take the limit $r\,{\to}\,0$ and find
\begin{equation}\label{eqn:derSr}
	\frac{d}{dx}S_w[i](x)\big|_{r=0}=\frac{1}{x-y_i}\left(\Sb_w[0](x)+\Sb_w[1](x)+S_{w_2\cdots w_n}[w_1](x)\right).
\end{equation}
Since the expressions here and in the following are convergent and well-defined for all $r\,{\geq}\,0$, it is possible to carry the dependence of $r$ until the very end (i.e.~until after the calculation of the boundary values $C_0$ and $C_1$) and then take it to zero, but it simplifies calculations to do so here already. All results for the matrices $e_0$ and $e_1$ discussed below will be written in the limit $r\,{\to}\,0$ already.

Having found the derivative of the integrals $S_w[i](x)$ in \eqn{eqn:derSr}, we can write down KZ equations for vectors of these functions. Before doing so for the general case, we will consider three examples to illustrate the procedure. 

\begin{example}\label{ex:0}
The simplest example is the insertion of the polylogarithms corresponding to the empty word $w\,{=}\,e$, i.e.~$L_e\,{=}\,1$, which describes the flat-space relation for the Veneziano amplitude described in \rcites{Broedel:2013aza,BK,Kaderli}: 
The vector $F_e(x)\,{=}\,(\Sb_e[0](x),\Sb_e[1](x))^T$ is a solution to the KZ equation \eqref{eqn:kzf} with the matrices
\begin{equation}
	e_0=\left(\begin{array}{cc}s&s\\0&0\end{array}\right),\qquad e_1=\left(\begin{array}{cc}0&0\\t&t\end{array}\right).
\end{equation}
Thus, the Drinfeld recursion \eqref{eqn:Drinrec} generates the four-point flat-space integrals (i.e.~the Veneziano amplitude) via
\begin{align}
\begin{split}
	C_1&=\left(\begin{array}{c}s\,J_e^{s,t+1}\\1\,{-}\,t\,J_e^{s+1,t}\end{array}\right)=\left(\begin{array}{c}\frac{\Gamma(s+1)\Gamma(1+t)}{\Gamma(1+s+t)}\\1{-}\frac{\Gamma(s+1)\Gamma(1+t)}{\Gamma(1+s+t)}\end{array}\right)\\
	&=\Phi(e_0,e_1)\cdot C_0=(\mathbb{I}-\zeta(2)[e_0,e_1]+\ldots)\cdot \left(\begin{array}{c}1\\0\end{array}\right).
\end{split}
\end{align}
\end{example}
\begin{example} \label{ex:1}
We want to generate the integral $J_0$ through the Drinfeld associator. For the insertion of the logarithm $L_0(x)\,{=}\,\log(x)$ into the Euler beta function integral, we consider the corresponding deformed integrals $S_0[0](x)$ and $S_0[1](x)$. Since their derivatives involve the integral $S_0[2](x)$, we can use relation \eqref{eqn:Srel} which will bring $S_e[0](x)$ into the game, whose derivative contains $S_e[1](x)$, which we then also need to include in the vector for the KZ equation \eqref{eqn:kzf}.

Thus, we define the vector 
{\small
\begin{equation}
	F_0(x)=\left(\begin{array}{c} \Sb_0[0](x) \\ \Sb_0[1](x) \\ \Sb_e[0](x) \\ \Sb_e[1](x)\end{array}\right),
\end{equation}
}
\!\!and using \eqn{eqn:derSr} we find that $F_0(x)$ is a solution to the KZ equation \eqref{eqn:kzf} for the $4\times4$ matrices (as mentioned before, $r$ has been set to zero in these matrices already)
{\small
\begin{equation}\label{eqn:eEx1}
	e_0=\left(\begin{array}{cccc} s & s & 1 & 0 \\ 0 & 0 & 0 & 0 \\ 0 & 0 & s & s \\ 0 & 0 & 0 & 0 \end{array}\right),\qquad 
	e_1=\left(\begin{array}{cccc} 0 & 0 & 0 & 0 \\ t & t & \tfrac{t}{s} & 0 \\ 0 & 0 & 0 & 0 \\ 0 & 0 & t & t \end{array}\right).
\end{equation}
}
\!\!\!The regularized boundary values $C_0,\, C_1$ from \eqn{eqn:boundval} are related via the Drinfeld recursion \eqref{eqn:Drinrec} with the matrices from \eqn{eqn:eEx1}.
Using the matrix exponentials $x^{{-}e_0}\,{=}\,\exp({-}e_0\log x)$ and $(1\,{-}\,x)^{{-}e_1}\,{=}\,\exp({-}e_1\log(1{-} x))$, we find the boundary values
\begin{equation}
	C_0=\left(\begin{array}{c}-\tfrac{1}{s} \\ 0 \\ 1 \\ 0 \end{array}\right),\qquad	C_1=\left(\begin{array}{c}s\,J_0^{s,t+1} \\ -t\,J_0^{s+1,t} \\ \tfrac{\Gamma(s+1)\Gamma(t+1)}{\Gamma(s+t+1)} \\ 1-\tfrac{\Gamma(s+1)\Gamma(t+1)}{\Gamma(s+t+1)}\end{array}\right).
\end{equation}
Focusing on the first component of $C_1$, the Drinfeld recursion yields an all-order expansion of the integral $s\,J_0^{s,t+1}$ in $s$ and $t$ with MZVs as coefficients, the first terms reading
\begin{align}
	s\,J_0^{s,t+1}&=C_1^{(1)}=(1,0,0,0)\cdot C_1=(1,0,0,0)\cdot \Phi(e_0,e_1)\cdot C_0\notag\\
	&=(1,0,0,0)\cdot\left(\mathbb{I}-\zeta(2)[e_0,e_1]-\zeta(3)[e_0+e_1,[e_0,e_1]]+\ldots\right)\cdot(-1/s,0,1,0)^T\notag\\
	&=-\frac{1}{s}+s \left(\zeta (3) t-\frac{1}{4} \zeta (4) t^2+\cO\!\left(t^3\right)\right)\notag\\
	&\quad+s^2 \left(-2\zeta (4) t+(4 \zeta (5)-2 \zeta (2) \zeta (3)) t^2+\cO\!\left(t^3\right)\right)+\cO\!\left(s^3\right).
\end{align}
\end{example}
\begin{example} \label{ex:2}
We consider the insertion of the trilogarithm $L_{001}(x)$ into the Euler beta integral, i.e.~we want to find a relation for the integral $J_{001}$. This is one of the $J$-integrals relevant for the first curvature correction to the Veneziano amplitude. Again we use the vector of functions $\Sb_w[i](x)$, $i\,{=}\,0,1$, with $w\in\{001,01,1,e\}$, i.e.
{\small
\begin{equation}
	F_{001}(x)=\left(\begin{array}{c}\Sb_{001}[0](x)\\\Sb_{001}[1](x)\\\Sb_{01}[0](x)\\\Sb_{01}[1](x)\\\Sb_{1}[0](x)\\\Sb_{1}[1](x)\\\Sb_{e}[0](x)\\\Sb_{e}[1](x)\end{array}\right).
\end{equation}
}
\!\!Using the relation \eqref{eqn:derSr}, we find that $F_{001}(x)$ solves the KZ equation \eqref{eqn:kzf} with matrices (at $r=0$)
{\small
\begin{equation}\label{eqn:ex2mat}
	e_0=\left(\begin{array}{cccccccc}
		s&s&1&0&0&0&0&0\\
		0&0&0&0&0&0&0&0\\
		0&0&s&s&1&0&0&0\\
		0&0&0&0&0&0&0&0\\
		0&0&0&0&s&s&0&\frac{s}{t}\\
		0&0&0&0&0&0&0&0\\
		0&0&0&0&0&0&s&s\\
		0&0&0&0&0&0&0&0
	\end{array}\right),\qquad
	e_1=\left(\begin{array}{cccccccc}
		0&0&0&0&0&0&0&0\\
		t&t&\frac{t}{s}&0&0&0&0&0\\
		0&0&0&0&0&0&0&0\\
		0&0&t&t&\frac{t}{s}&0&0&0\\
		0&0&0&0&0&0&0&0\\
		0&0&0&0&t&t&0&1\\
		0&0&0&0&0&0&0&0\\
		0&0&0&0&0&0&t&t
	\end{array}\right).
\end{equation}
}
\!\!The regularized boundary values \eqref{eqn:boundval} of $F_{001}(x)$ at $x\,{=}\,0,1$ are found to be
{\small
\begin{equation}\label{eqn:C1-001}
	C_0=\left(\begin{array}{c}0\\0\\0\\0\\0\\0\\1\\0\end{array}\right),\qquad 
	C_1=\left(\begin{array}{c}s\,J_{001}^{s,t+1}\\{-}s\,J_{001}^{s,t+1}{-}J_{01}^{s,t+1}{-}\zeta(3)\\s\,J_{01}^{s,t+1}\\{-}s\,J_{01}^{s,t+1}{-}J_{1}^{s,t+1}{-}\zeta(2)\\s\,J_{1}^{s,t+1}\\{-}s\,J_{s,t+1}{+}\frac{s}{t}J_e^{s,t+1}{-}\frac{1}{t}\\\frac{\Gamma(s+1)\Gamma(1+t)}{\Gamma(1+s+t)}\\1{-}\frac{\Gamma(s+1)\Gamma(1+t)}{\Gamma(1+s+t)}\end{array}\right).
\end{equation}
}
\!\!This allows us to obtain the integral $J_{001}^{s,t+1}$ as an infinite series in the variables $s,\,t$ using the Drinfeld recursion \eqref{eqn:Drinrec} and yields the expansion
\begin{align}
	J_{001}^{s,t+1}&=-\zeta (4)+(3 \zeta (5){-}\zeta (2) \zeta (3)) t+\!\left(\!\frac{\zeta (3)^2}{2}{-}\frac{89 \zeta (6)}{48}\!\right)\! t^2+\cO\left(t^3\right)\notag\\
	&\quad+s\! \left(\!\zeta (5)+\!\left(\!\frac{\zeta (6)}{3}{-}\frac{\zeta (3)^2}{2}\!\right)\! t+\!\left(\!2\zeta (3) \zeta (4){-}9 \zeta (5) \zeta (2){+}\frac{205 \zeta (7)}{16}\!\right)\! t^2+\cO\!\left(t^3\right)\!\right)\notag\\
	&\quad+s^2 \!\left(\!-\zeta (6)+(12 \zeta (7)-7 \zeta (2) \zeta (5)) t+\cO\left(t^3\right)\!\right)\!+\cO\left(s^3\right).
\end{align}
\end{example}
Results for all other weight three insertions are given in \appref{app:explicit3}.

%%%%%%%%%%%%%%%%%%%%%%%%%%%%%%%%%
\paragraph{General case.}
Let us consider the scenario of an arbitrary word $w\,{=}\,w_1\cdots w_n\,{\in}\,\{0,1\}^\times$. The vector $F_w(x)$ is then the $2(n{+}1)$-dimensional vector
\begin{equation}
	F_w(x)=\left(\begin{array}{c}\Sb_w[0](x)\\\Sb_w[1](x)\\\Sb_{w_2\cdots w_n}[0](x)\\\Sb_{w_2\cdots w_n}[1](x)\\\vdots\\\Sb_{w_n}[0](x)\\\Sb_{w_n}[1](x)\\\Sb_{e}[0](x)\\\Sb_{e}[1](x)\end{array}\right).
\end{equation}
As before, relation \eqref{eqn:derSr} can be used to show that $F_w(x)$ is a solution to the KZ equation \eqref{eqn:kzf}. The structure of the matrices $e_0$ and $e_1$ is as follows:

We consider a subblock of the vector $F_w$ for some word $w'$ which fulfills $w\,{=}\,\cdots w'$. We can have two different cases, depending on which letter appears in $w$ just before $w'$, i.e.~if $w\,{=}\,\cdots w_iw'$, whether $w_i\,{=}\,0$ or $w_i\,{=}\,1$.
\begin{itemize}
	\item {\boldmath{$w_i\,{=}\,0\,{:}$}} We consider the subblock of $F_w$ reading
	\begin{equation}
		F_w(x)=\left(\begin{array}{c}\vdots\\\Sb_{0w'}[0](x)\\\Sb_{0w'}[1](x)\\\Sb_{w'}[0](x)\\\Sb_{w'}[1](x)\\\vdots\end{array}\right).
	\end{equation}
	Then relation \eqref{eqn:derSr} shows that the matrices $e_0$ and $e_1$ have the corresponding blocks (again we have set $r\,{=}\,0$ already)
	\begin{equation}\label{eqn:mat1}
		e_0=\left(\begin{array}{cccccc}
			\ddots & \vdots & \vdots & \vdots & \vdots & \iddots \\
			\hdots & s & s & 1 & 0 & \hdots\\
			\hdots & 0 & 0 & 0 & 0 & \hdots\\
			\hdots & 0 & 0 & s & s & \hdots\\
			\hdots & 0 & 0 & 0 & 0 & \hdots\\
			\iddots & \vdots & \vdots & \vdots & \vdots & \ddots
		\end{array}\right),\qquad
		e_1=\left(\begin{array}{cccccc}
			\ddots & \vdots & \vdots & \vdots & \vdots & \iddots \\
			\hdots & 0 & 0 & 0 & 0 & \hdots\\
			\hdots & t & t & \frac{t}{s} & 0 & \hdots\\
			\hdots & 0 & 0 & 0 & 0 & \hdots\\
			\hdots & 0 & 0 & t & t & \hdots\\
			\iddots & \vdots & \vdots & \vdots & \vdots & \ddots
		\end{array}\right).
	\end{equation}
	\item {\boldmath{$w_i\,{=}\,1\,{:}$}} In this case, we take the subblock
	\begin{equation}
		F_w(x)=\left(\begin{array}{c}\vdots\\\Sb_{1w'}[0](x)\\\Sb_{1w'}[1](x)\\\Sb_{w'}[0](x)\\\Sb_{w'}[1](x)\\\vdots\end{array}\right),
	\end{equation}
	for which we have the block structure (at $r\,{=}\,0$)
	\begin{equation}\label{eqn:mat2}
		e_0=\left(\begin{array}{cccccc}
			\ddots & \vdots & \vdots & \vdots & \vdots & \iddots \\
			\hdots & s & s & 0 & \frac{s}{t} & \hdots\\
			\hdots & 0 & 0 & 0 & 0 & \hdots\\
			\hdots & 0 & 0 & s & s & \hdots\\
			\hdots & 0 & 0 & 0 & 0 & \hdots\\
			\iddots & \vdots & \vdots & \vdots & \vdots & \ddots
		\end{array}\right),\qquad
		e_1=\left(\begin{array}{cccccc}
			\ddots & \vdots & \vdots & \vdots & \vdots & \iddots \\
			\hdots & 0 & 0 & 0 & 0 & \hdots\\
			\hdots & t & t & 0 & 1 & \hdots\\
			\hdots & 0 & 0 & 0 & 0 & \hdots\\
			\hdots & 0 & 0 & t & t & \hdots\\
			\iddots & \vdots & \vdots & \vdots & \vdots & \ddots
		\end{array}\right).
	\end{equation}
\end{itemize}
All entries of $e_0$ and $e_1$ not covered by these considerations vanish (c.f.~\exref{ex:1} and \exref{ex:2}).

The regularized boundary values\footnote{Whether we take $r\,{\to}\,0$ before or after taking $x$ to 0 or 1 does not change the result.} $C_0$ and $C_1$ from \eqn{eqn:boundval} can then be calculated using the matrix exponential. For $C_1$ we find
\begin{equation}
	C_1=\left(\begin{array}{c}s\,J_w^{s,t+1}\\\cdots\\s\,J_{w_2\cdots w_n}^{s,t+1}\\\cdots\\\vdots\\s\,J_{w_n}^{s,t+1}\\\cdots\\s\,J_{e}^{s,t+1}\\1-t\,J_{e}^{s+1,t}\end{array}\right),
\end{equation}
where every even component $C_1^{(2i)}$, $i\,{=}\,1,\ldots,n$, depends on the combination of 0's and 1's in the word $w$ (and thus is here only denoted by ``$\cdots$'') and for the odd components we have $C_1^{(2i-1)}\,{=}\,s\,J_{w_i\cdots w_n}^{s,t+1}$. For examples, the reader can find all weight three results in \appref{app:explicit3}.

For $C_0$ we have to distinguish whether the word $w$ ends on $0$ or $1$. If $w$ ends with 1, $C_0$ simply reads
\begin{equation}\label{eqn:c0}
	C_0=\left(0,\cdots,0,1,0\right)^T,
\end{equation}
so we only find a 1 in the second to last entry (as the bottom two-dimensional subblock yields the flat-space string-amplitude recursion from \rcite{Broedel:2013aza,BK,Kaderli} discussed in \exref{ex:0}) and all other entries are 0.

If on the other hand $w$ ends with $k$ 0's, i.e.~$w\,{=}\,\cdots 10^k$, we find (c.f.~examples in \appref{app:explicit3})
\begin{equation}\label{eqn:c00}
	C_0=\left(0,\cdots,0,0,\frac{(-1)^k}{s^k},0,\cdots,\frac{1}{s^2},0,-\frac{1}{s},0,1,0\right)^T.
\end{equation}

The Drinfeld recursion \eqref{eqn:Drinrec} then relates the boundary values and provides all-order expansions of the integrals $J_w^{s,t+1}$ for any word $w\,{\in}\,\{0,1\}^\times$. While already shown in \rcite{Alday:2024yax}, the recursive techniques presented here give another proof that the integrals $J_w^{s,t}$ admit an expansion in $s$ and $t$ with coefficients valued in the ring $\cZ$ of MZVs over $\mathbb{Q}$.

Note that the recursion as it is phrased here will actually yield $s\,J_w^{s,t+1}$ as components of $C_1$. If one wants to obtain $J_w^{s,t}$ as studied in \rcite{Alday:2025bjp}, one can use the shift relations of \rcite{Alday:2025bjp} shown here in \eqn{eqn:shiftrel}. For example one finds the relation

\begin{align}\label{eqn:J001}
\begin{split}
	J_{001}^{s,t}&=\frac{1}{t}\left((s+t)J_{001}^{s,t+1}+J_{01}^{s,t+1}\right)\\
	&=\frac{1}{s t}\left(s+t,0,1,0,0,0,0,0\right)\cdot C_1\\
	&=\frac{1}{s t}\left(s+t,0,1,0,0,0,0,0\right)\cdot \Phi(e_0,e_1)\cdot C_0\\
	&=\frac{1}{s t}\left(s+t,0,1,0,0,0,0,0\right)\cdot \Phi(e_0,e_1)\cdot (0,0,0,0,0,0,1,0)^T,
\end{split}
\end{align}
with the vectors $C_0,\,C_1$ from \eqn{eqn:C1-001} and matrices $e_0,\,e_1$ from \eqn{eqn:ex2mat} of \exref{ex:2}. This yields a simple relation that generates the integral $J_{001}^{s,t}$ from the Drinfeld associator.

Corresponding relations for all integrals $J_w^{s,t}$ with $w$ of weight three can be found in \appref{app:explicit3}.

%%%%%%%%%%%%%%%%%%%%%%%%%%%%%%%%%
\paragraph{Higher-point relation.}
Although the function space necessary to describe higher-point string amplitudes on AdS backgrounds is not yet known\footnote{In \rcite{Alday:2025bjp} it was suggested that insertions of polylogarithms in two variables (see e.g.~\rcites{DelDuca:2016lad,Broedel:2016kls,Frost:2023stm,Frost:2025lre}) might be needed, which could be a different structure than the one studied in the following. Nevertheless, it will be mathematically interesting to investigate the structure of iterated integrals with polylogarithmic insertions, and one might hope to identify this structure in certain limits of higher-point AdS amplitudes in the future.}, we can extend the structure discussed before to iterated integrals similar to higher-point open-string integrals with extra insertions of polylogarithms in each integration. Thus, we are interested in integrals of the type\footnote{Note, that for simplicity we shifted the exponents $s_{ij}{-}1\mapsto s_{ij}$, compared to \eqn{eqn:Jint}.}
\begin{equation}\label{eqn:JN}
	J_{w_3,\ldots,w_k}^{\{s_{ij}\}}=\int_{0\leq x_k\leq\cdots\leq x_3\leq1}\prod_{j=3}^{k}dx_j\,L_{w_j}(x_j)\prod_{\substack{0\leq i<j\leq k\\i,j\neq2}}(x_i-x_j)^{s_{ij}},
\end{equation}
which, for $k\,{\geq}\,3$, are Selberg-type iterated integrals with an extra polylogarithm insertion for each integral, corresponding to words $w_j\,{\in}\,\{0,1\}^\times$, $j\,{=}\,3,\ldots k$, and where we used $x_0\,{=}\,0,\,x_{1}\,{=}\,1$. Such iterated $J$-integrals can also appear as results of products of 1-fold $J$-integrals through the shuffle relation, e.g.~(for $w_1,w_2\,{\in}\,\{0,1\}^\times$ and parameters $s_1,t_1,s_2,t_2$ with positive real parts)
\begin{align}
\begin{split}
	J_{w_1}^{s_1,t_1}\,J_{w_2}^{s_2,t_2}&=\int_{0}^{1}dx_1\,x_1^{s_1}(1-x_1)^{t_1}L_{w_1}(x_1)\int_{0}^{1}dx_2\,x_2^{s_2}(1-x_2)^{t_2}L_{w_2}(x_2)\\
	&=\int_{0}^{1}dx_1\,x_1^{s_1}(1-x_1)^{t_1}L_{w_1}(x_1)\int_{0}^{x_1}dx_2\,x_2^{s_2}(1-x_2)^{t_2}L_{w_2}(x_2)\\
	&\quad+\int_{0}^{1}dx_2\,x_2^{s_2}(1-x_2)^{t_2}L_{w_2}(x_2)\int_{0}^{x_2}dx_1\,x_1^{s_1}(1-x_1)^{t_1}L_{w_1}(x_1)\\
	&=J_{w_1,w_2}^{s_1,t_1,s_2,t_2}+J_{w_2,w_1}^{s_2,t_2,s_1,t_1}.
\end{split}
\end{align}

In order to find an associator relation for these iterated $J$-integrals of \eqn{eqn:JN}, we again define interpolating functions
\begin{equation}\label{eqn:SN}
	S_{w_3,\ldots,w_k}[i_3,\ldots,i_k](x)=\int_{0\leq x_k\leq\cdots\leq x_3\leq x}\prod_{j=3}^{k}dx_j\frac{L_{w_j}(x_j)}{x_j-x_{i_j}}(x-x_j)^{r_j}\prod_{\substack{1\leq i<j\leq k\\i,j\neq2}}(x_i-x_j)^{s_{ij}},
\end{equation}
depending on an extra variable $x_{2}\,{=}\,x\,{\in}\,]0,1[$ appearing as the upper integration limit and in the integrand in the differences $(x-x_j)^{r_j}$. In order for the integrals in \eqns{eqn:JN}{eqn:SN} to converge, we assume\footnote{As mentioned in the four-point case already, the true region is larger and depends on the labels $i_1,\ldots,i_k$.} $\Re s_{ij},\Re r_j>0$.

We follow the same steps as in the four-point case earlier, i.e.~calculate the derivatives and $S$-integral relations as in \eqns{eqn:derS}{eqn:Srel} correspondingly for the integrals \eqref{eqn:SN}, to then write a Drinfeld recursion relating $k$-iterated $J$-integrals to $(k\,{+}\,1)$-iterated $J$-integrals via the Drinfeld associator. While the combinatorics get more cumbersome, the general structure remains the same and generalizes the $N$-point flat-space string-amplitude recursion from \rcites{Broedel:2013aza,BK,Kaderli,Baune:2024uwj}.

To illustrate this, we will here consider one example of a two-fold integral with one polylogarithmic insertion.

\begin{example}\label{ex:3}
	We consider the case $k\,{=}\,4$ with $w_3\,{=}\,1$, $w_4\,{=}\,e$, as for higher cases the matrices quickly become large, but the structure remains essentially the same. The vector for the KZ equation will be
	{\small
	\begin{equation}
		F_{1,e}(x)=\left(\begin{array}{c}
			\Sb_{1,e}[0,0](x)\\
			\Sb_{1,e}[0,1](x)\\
			\Sb_{1,e}[0,3](x)\\
			\Sb_{1,e}[1,0](x)\\
			\Sb_{1,e}[1,1](x)\\
			\Sb_{1,e}[1,3](x)\\
			\Sb_{e,e}[0,0](x)\\
			\Sb_{e,e}[0,1](x)\\
			\Sb_{e,e}[0,3](x)\\
			\Sb_{e,e}[1,0](x)\\
			\Sb_{e,e}[1,1](x)\\
			\Sb_{e,e}[1,3](x)\\
			\end{array}\right).
	\end{equation}
	}
	\!\!\!\!where we used the shorthand $\Sb_{w_3,w_4}[i_3,i_4](x)\,{=}\,s_{3i_1}s_{4i_2}\,S_{w_3,w_4}[i_3,i_4](x)$ and again $x_0\,{=}\,0,\,x_{1}\,{=}1,\,x_2\,{=}\,x$.
	
	Now we calculate the derivatives, e.g.
	\begin{equation}
		\frac{d}{dx}\Sb_{1,e}[0,0](x)=\frac{s_{03}s_{04}}{x}(r_3(S_{1,e}[0,0](x)\,{-}\,S_{1,e}[2,0](x))\,{+}\,r_4(S_{1,e}[0,0](x)\,{-}\,S_{1,e}[0,2](x))),
	\end{equation}
	and derive relations analogous to \eqn{eqn:Srel}, e.g.~for $i_4\neq3$
	\begin{align}
		0&=\!\!\int_{0}^x\!\!dx_3\frac{d}{dx_3}L_1(x_3)(x{-}x_3)^{r_3}\!\!\int_{0}^{x_3}\!\!\frac{dx_4}{x_4{-}x_{i_4}}(x{-}x_4)^{r_4}x_3^{s_{03}}(1{-}x_3)^{s_{13}}x_4^{s_{04}}(1{-}x_4)^{s_{14}}(x_3{-}x_4)^{s_{34}}\notag\\
		&=r_3\,S_{1,e}[2,i_4](x)+s_{03}\,S_{1,e}[0,i_4](x)+s_{13}\,S_{1,e}[1,i_4](x)+s_{34}\,S_{1,e}[4,i_4](x)+S_{e,e}[1,i_4](x)\notag\\
		&=r_3\,S_{1,e}[2,i_4](x)+s_{03}\,S_{1,e}[0,i_4](x)+s_{13}\,S_{1,e}[1,i_4](x)\notag\\
		&\quad-s_{34}\,S_{1,e}[i_4,3](x)+s_{34}\,S_{1,e}[i_4,i_4](x)+S_{e,e}[1,i_4](x),
	\end{align}
	where we used partial fractioning to get to the last line.
	Using these results together (to eliminate appearances of index 2 in the results of the derivatives), we find that $F_{1,e}(x)$ solves the KZ equation \eqref{eqn:kzf}. The $12\times12$ matrices $e_0$ and $e_1$ are shown in \appref{app:5pt}.
	The boundary values read
	\begin{subequations}
	\begin{align}
		C_0&=(0,0,0,0,0,0,1,0,-1,0,0,0)^T\frac{s_{03}\,\Gamma(s_{04}{+}1) \Gamma(s_{34}{+}1)}{(s_{03}{+}s_{04}{+}s_{34}) \Gamma (s_{04}{+}s_{34}{+}1)},\\
		C_1&=(s_{03}s_{04}J_{1,e}^{\{s_{03}-1,s_{04}-1,s_{13},s_{14},s_{34}\}},\ldots)^T,
	\end{align}
	\end{subequations}
	so that the Drinfeld recursion \eqref{eqn:Drinrec} relates depth-1 and depth-2 $J$-integrals among each other, and e.g.~the low-energy expansion of the integral $J_{1,e}$ can readily be generated through the Drinfeld associator using the matrix representations for $e_0$ and $e_1$ provided in \appref{app:5pt}.
\end{example}

In general, this formalism provides a relation that allows to generate an $n$-fold iterated $J$-integral (in $C_1$) from a set of $(n{-}1)$-fold iterated $J$-integrals (from $C_0$) through the Drinfeld associator. Starting with the four-point cases discussed earlier (1-fold integration), this implies that the iterated $J$-integrals defined in \eqn{eqn:JN} admit expansions in the parameters $s_{ij}$ with coefficients valued in $\cZ$.

\subsection{Closed-string building blocks and the Deligne associator}\label{sec:closedrec}
Similarly to \rcite{Baune:2024uwj}, where the open-string recursion was lifted to a closed-string recursion for the flat-space amplitudes, we also want to lift the results of the open-string results of the previous subsection to the closed-string case, to find associator relations for the $I$-integrals from \eqn{eqn:Iint}.

Since the $I$-integrals are the single-valued images of the $J$-integrals, i.e.~$I_w^{s,t}\,{=}\,\sv(J_w^{s,t})$ \cite{Alday:2025bjp}, we merely need to map all objects of the formalism of \subsecref{sec:openrec} to their single-valued analogues. We find the boundary values\footnote{Since $\sv(1)\,{=}\,1$ and $C_0$ has the simple form of \eqref{eqn:c0} and \eqref{eqn:c00}, we find $\cC_0=\sv(C_0)=C_0$.}$^,$\footnote{Note here, that generally $\sv(J_w^{s,t+1})\neq I_w^{s,t+1}$, but one has to use the shift relations \eqref{eqn:shiftrel} to first convert $J_w^{s,t+1}$ into a linear combination of integrals $J_{w'}^{s,t}$ to then use $\sv(J_{w'}^{s,t})=I_{w'}^{s,t}$ (which can be converted into $I_{w''}^{s,t+1}$ using the single-valued shift relations \eqref{eqn:svshiftrel} if desired). E.g.
	\begin{equation*}
		\sv(J_{001}^{s,t+1})=\frac{t}{st}I_{001}^{s,t}-\frac{t}{(st)^2}I_{01}^{s,t}+\frac{t}{(st)^3}I_1^{s,t}+\frac{1}{(st)^3}\left(1-\frac{t}{st}\right)I_e^{s,t}.
	\end{equation*}
	}
\begin{equation}
	\cC_0=\sv(C_0)=C_0,\qquad \cC_1=\sv(C_1),
\end{equation} 
and the Drinfeld associator $\Phi$ becomes its single-valued analogue, the Deligne associator from \eqn{eqn:associators}, $\Phi^{\sv}\,{=}\,\sv(\Phi)$.
Thus, we have the Deligne recursion \eqref{eqn:Delinrec} for the closed-string building blocks with the same matrices $e_0$ and $e_1$ as in the open-string case.

Since we find the $I$-integrals as components of $\cC_1$, and since $\Phi^{\sv}$ only contains svMZVs as coefficients, this is another proof (complementing the proof in \rcite{Alday:2023jdk}) that the expansion in $s$ and $t$ of the integrals $I_w^{s,t}$ has coefficients valued in $\cZ^{\sv}$. Furthermore, this Deligne recursion~\eqref{eqn:Delinrec} provides an all-order expression in $s$ and $t$ of the $I$-integrals.

Let us illustrate this with one example for the closed-string case.

\begin{example}\label{ex:4}
	We want to generate an expansion of the integral $I_{001}^{s,t}\,{=}\,\sv(J_{001}^{s,t})$ using the Deligne associator.
	Using the single-valued version of \eqn{eqn:J001} we find
	\begin{align}
	\begin{split}
		I_{001}^{s,t}&=\frac{1}{s t}\left(s+t,0,1,0,0,0,0,0\right)\cdot \cC_1\\
		&=\frac{1}{s t}\left(s+t,0,1,0,0,0,0,0\right)\cdot \Phi^{\sv}(e_0,e_1)\cdot C_0\\
		&=\frac{1}{s t}\left(s+t,0,1,0,0,0,0,0\right)\cdot \Phi^{\sv}(e_0,e_1)\cdot (0,0,0,0,0,0,1,0)^T.
	\end{split}
	\end{align}
\end{example}
All other relations for the integrals $I_w^{s,t}$ with $w$ of weight three can be found through \appref{app:explicit3}, where one simply has to replace $\Phi$ by $\Phi^{\sv}$ in the Drinfeld relations given there for the corresponding integrals $J_w^{s,t}$.

%%%%%%%%%%%%%%%%%%%%%%%%%%%%%%%%%
\paragraph{Higher-point relation.}
For the iterated $J$-integrals from \eqn{eqn:JN} discussed at the end of \subsecref{sec:openrec}, we saw that the four-point structure generalizes to iterated integrals, reminiscent of flat-space open-string integrals with polylogarithm insertion. This is also true in the closed-string case: Using the methods of \rcites{Brown:2018omk,Brown:2019wna}, the single-valued images of the results for the higher-point cases at the end of \subsecref{sec:openrec} can be lifted to the closed-string setting to obtain Deligne recursion relations for iterated $I$-integrals of the form
\begin{equation}\label{eqn:IN}
	I_{w_1,\ldots,w_k}^{\{s_{ij}\}}=\prod_{j=3}^{k}\int_{\mathbb{C}}d^2z_j\,\cL_{w_j}(z_j)\prod_{\substack{0\leq i<j\leq k\\i,j\neq2}}|z_i-z_j|^{2s_{ij}},
\end{equation}
for $w_1,\ldots,w_k\,{\in}\,\{0,1\}^\times$ and using $z_0\,{=}\,0,\,z_1\,{=}\,1$. Thus, these integrals can be generated from the Deligne associator and starting from the four-point result, this recursively shows that the iterated $I$-integrals from \eqn{eqn:IN} admit expansions in $s$ and $t$ with coefficients in the ring $\cZsv$ of svMZVs.

%%%%%%%%%%%%%%%%%%%%%%%%%%%%%%%%%%%%%%%
\section{Outlook}
\label{sec:conclusions}
Having identified the associators for the building blocks of AdS amplitudes for open and closed strings, this opens up further research directions, some of which we highlight here.
\begin{itemize}
\item The methods described in this paper with integrations over one-forms containing logarithms and closing under taking derivatives might be a starting point to investigate the (co-)homologies for AdS amplitudes more closely, using the flat-space methods from \rcites{Mizera:2016jhj,Mizera:2017cqs,Mizera:2017rqa,Kaderli}. This could also allow to find the intersection numbers related to the AdS-KLT relations found in \rcite{Alday:2025bjp}.
\item From a structural perspective, one might wonder how this construction of Selberg integrals with polylogarithm insertions translates to the elliptic or higher-genus cases. One could consider integrals similar to one-loop string amplitudes but with additional insertions of elliptic polylogarithms. Recursive relations for flat-space one-loop integrals have already been studied in \rcites{Mafra:2019ddf,Mafra:2019xms,BK,Broedel:2020tmd}, and one could generalize these structures to integrals with elliptic polylogarithm insertions. Whether such functions would actually be related to one-loop string amplitudes on AdS backgrounds remains to be settled in the future.
\item Finally, the relations found here provide insights into the building blocks of AdS string amplitudes, but it would be interesting to rewrite such relations for precise combinations of $J$- and $I$-integrals as they appear in the physical theories. This might yield deeper understanding of the structure of these scattering amplitudes.
\end{itemize}
%

%%%%%%%%%%%%%%%%%%%%%%%%%%%%%%%%%%
%%%%%%%%%%%%%%%%%%%%%%%%%%%%%%%%%%
\acknowledgments

The author is grateful to Johannes Broedel for various discussions, work on related projects and comments on a draft of this article and to Kiarash Naderi for discussion.\\
The work of the author is partially supported by the Swiss National Science Foundation through the NCCR SwissMAP.

%\paragraph{Note added.} This is also a good position for notes added after the paper has been written.

%%%%%%%%%%%%%%%%%%%%%%%%%%%%%%%%%%%%%%%%%%%%%%%%%%%%%%%%%%%%%%%%%%%%%%
%%%%%%%%%%%%%%%%%%%%%%%%%%%%%%%%%%%%%%%%%%%%%%%%%%%%%%%%%%%%%%%%%%%%%%

\appendix
%%%%%%%%%%%%%%%%%%%%%%%%%%%%%%%%%%%%%%%%%%%%%%%%%%%%%%%%%%%%%%%%%%%%%

\section{Polylogarithms, multiple zeta values and associators}
\label{app:polylogs}
In this appendix we will review the basics about polylogarithm, multiple zeta values and their single-valued analogs and provide the background knowledge about the Drinfeld and Deligne associators. We will follow the conventions of \rcite{Alday:2025bjp} and as (up to conventions) this material is widely used, we will merely reference useful overview articles on the topic, namely \rcites{Brown:2013qva,Brownhyperlogs,Brown:2013gia}.

\subsection{Polylogarithms and multiple zeta values}\label{sec:MPLs}
One generalization of the logarithm are the \emph{classical polylogarithms}
\begin{equation}
	\Li_n(x)=\sum_{k=1}^\infty\frac{x^k}{k^n}.
\end{equation}
For a word\footnote{The set $\{0,1\}^\times$ contains all words made from the letters 0 and 1 including the empty word $e$.} $w\,{=}\,w_3\cdots w_n\,{\in}\,\{0,1\}^{\times}$, the classical polylogarithms can be generalized to \emph{multiple polylogarithms} (MPLs)
\begin{align}
\begin{split}
	L_{w}(x)&=\int_0^x\frac{dx_3}{x_3-x_{w_3}}L_{w_4\cdots w_n}(x_3)\\
	&=\prod_{j=3}^{n}\int_0^{x_{j-1}}\frac{dx_j}{x_j-x_{w_j}},
\end{split}
\end{align}
where we used the notation $x_0\,{=}\,0,\,x_1\,{=}\,1,\,x_2\,{=}\,x$.

Classical polylogarithms evaluated at $x\,{=}\,1$ yield special values of the Riemann zeta function
\begin{equation}
	\Li_n(1)=\zeta(n),
\end{equation}
and this notion is then generalized to define \emph{multiple zeta values} (MZVs) as the special values at $x\,{=}\,1$ of MPLs,
\begin{equation}
	L_{w}=(-1)^{\# w}\zeta_w.
\end{equation}
where $\# w$ is the number of 1's in the word $w\,{\in}\,\{0,1\}^\times$. The MZVs $\zeta_w$ in the binary notation (i.e.~$w\,{\in}\,\{0,1\}^\times$) can be transformed into the collapsed notation according to
\begin{equation}
	\zeta_{0^{n_1-1}1\cdots0^{n_k-1}}=\zeta(n_1,\ldots,n_k)=\sum_{j_1>\ldots>j_k>0}\prod_{i=1}^k j_i^{-n_i}.
\end{equation}
As an inherent property of iterated integrals, the MZVs obey a shuffle relation
\begin{equation}\label{eqn:shuffle}
	\zeta_{w_1}\,\zeta_{w_2}=\sum_{w\in w_1\shuffle w_2}\zeta_w,
\end{equation}
for $w_1,\,w_2\,{\in}\,\{0,1\}^\times$ and where the shuffle product $w_1\,{\shuffle}\, w_2$ is the set of all words from the letters of $w_1$ and $w_2$ while keeping the respective orders intact.
This gives rise to the ring $\cZ$ of MZVs over $\mathbb{Q}$.

The generating function of polylogarithms
\begin{equation}\label{eqn:polygen}
	L(x)=\sum_{w\in\{e_0,e_1\}^\times}w\,L_w(x)
\end{equation}
is a solution to the \emph{Knizhnik--Zamolodchikov equation}
\begin{equation}
	\label{eqn:kzeqpoly}
	\frac{d}{dx}L(x)=\left(\frac{e_0}{x}+\frac{e_1}{x-1}\right)L(x).
\end{equation}
This equation tells us the behavior of $L(x)$ near $x\,{=}\,0$, which is $\frac{d}{dx}L(x)\,{\sim}\,\frac{e_0}{x}L(x)$, implying $L(x)\,{\sim}\, x^{e_0}$. By symmetry under $x\,{\mapsto}\,1\,{-}\,x$ of the above equation, there exists another solution $L'(x)$ with behavior $L'(x)\,{\sim}\,(1\,{-}\,x)^{e_1}$ near $x\,{=}\,1$. This solution is unique if one demands $\lim_{x\to1}(1\,{-}\,x)^{-e_1}L'(x)\,{=}\,1$. Since $L$ and $L'$ are both solutions to \eqn{eqn:kzeqpoly}, their combination\footnote{Note, that we omitted the dependence of $\Phi$ on the algebra generators $e_0$, $e_1$.}
\begin{equation}\label{eqn:phi}
	\Phi(e_0,e_1)=L'(x)^{-1}L(x)
\end{equation}
is independent of $x$. This combination $\Phi$ is called the \emph{Drinfeld associator}, which is the generating function of MZVs, as we can see by evaluating \eqn{eqn:phi} at $x\,{=}\,1$:
\begin{equation}
	\Phi(e_0,e_1)=L(1)=\sum_{w\in\{e_0,e_1\}^\times}(-1)^{\#w}w\,\zeta_w.
\end{equation}

\subsection{Single-valued polylogarithms and single-valued multiple zeta values}
\label{sec:svpolylogs}

Multiple polylogarithms as reviewed before are multi-valued functions inheriting branch cuts from the iterated integrals over simple poles. It is possible to combine MPLs in such a way that all branch cuts cancel. The resulting functions are single-valued on $\mathbb{C}\setminus\{0,1\}$ and thus called \emph{single-valued multiple polylogarithms} (svMPLs). The procedure on how to obtain svMPLs from MPLs was shown by Brown in \rcite{BrownSVMPL}. In the simplest case of the logarithm $L_0(x)\,{=}\,\log(x)$ one can get the single-valued analog by combining it with its complex conjugate to find the single-valued logarithm\footnote{Here and in the following we omit the explicit dependence on $\zb$ and write svMPLs simply as functions of a complex variable $z$.} $\cL_0(z)\,{=}\,\log(z)+\log(\zb)\,{=}\,\log(|z|^2)$. More involved cases are explicitly calculated in \rcite{Dixon:2012yy}, for example
\begin{align}\label{eqn:svmpls}
	\cL_{01}(z)&=L_{10}(z)\,{+}\,L_{01}(\zb)\,{+}\,L_1(z)\,L_0(\zb),\notag\\
	\cL_{001}(z)&=L_{001}(z)\,{+}\,L_{100}(\zb)\,{+}\,L_{00}(z)\,L_1(\zb)\,{+}\,L_0(z)\,L_{10}(\zb),\\
	\cL_{0011}(z)&=L_{0011}(z){+}L_{1100}(\zb){+}L_{001}(z)\,L_1(\zb){+}L_0(z)\,L_{110}(\zb){+}L_{00}(z)\,L_{11}(\zb){+}2\zeta(3)\,L_1(\zb).\notag
\end{align}
Formally, svMPLs can be found from solving the single-valued condition \cite{BrownSVMPL}
\begin{equation}\label{eqn:svcond}
	\Phi(e_0,e_1)^{-1}\,e_1\,\Phi(e_0,e_1)=\widetilde{\Phi}(e_0,e_1')\,e_1'\,\widetilde{\Phi}(e_0,e_1')^{-1},
\end{equation}
where $\widetilde{\phantom{m}}$ refers to reversal of words, for the letter $e_1'$ around $e_1$ recursively, giving to first orders
\begin{equation}\label{eqn:e1p}
	e_1'=e_1-2\zeta(3)[e_0+e_1,[e_1,[e_0,e_1]]]+\ldots\ .
\end{equation}
Then the generating series of svMPLs $\cL(z)$ is defined through the product of the generating series \eqref{eqn:polygen} of MPLs \cite{BrownSVMPL}
\begin{align}\label{eqn:cLgen}
\begin{split}
	\cL(z)&=L(e_0,e_1;z)\,\widetilde{L}(e_0,e_1';\zb)\\
	&=\sum_{w\in\{e_0,e_1\}^\times}w\,\cL_w(z).
\end{split}
\end{align}
Note, that the appearance of $\zeta(3)$ in the last line of \eqn{eqn:svmpls} is due to the corrections in $e_1'\,{=}\,e_1+\ldots$ from the condition \eqref{eqn:svcond}.

Analogously as one defined MZVs as the special values of MPLs, one defines \emph{single-valued multiple zeta values} (svMZVs) as the special values of svMPLs at $z\,{=}\,1$,
\begin{equation}
	\cL_w(1)=(-1)^{\# w}\zetasv_w.
\end{equation}
Just as the MZVs, the svMZVs fulfill the shuffle relations \eqref{eqn:shuffle}, giving rise to the ring~$\cZ^{\sv}$ of svMZVs over $\mathbb{Q}$.
This yields a map $\sv{:}\ \cZ\to\cZsv$ mapping a MZV $\zeta_w$ to its single-valued analog $\zetasv_w\,{=}\,\sv(\zeta_w)$~\cite{Brown:2013gia}.

The generating series of svMPLs defined in \eqn{eqn:cLgen} fulfills the holomorphic equation \eqref{eqn:kzeqpoly} in $z$ as well as the antiholomorphic equation~\cite{Brown:2013gia}
\begin{equation}
	\frac{\partial}{\partial \zb}\cL(z)=\cL(z)\left(\frac{e_0}{\zb}+\frac{e_1'}{\zb-1}\right).
\end{equation}
As described in \rcite{Brown:2013gia}, the corresponding associator for this setup is the \emph{Deligne associator} $\Phi^{\sv}$, defined through the Drinfeld associator $\Phi$ as
\begin{equation}
	\Phi^{\sv}(e_0,e_1)=\Phi(e_0,e_1)\,\widetilde{\Phi}(e_0,e_1').
\end{equation}
The Deligne associator is the single-valued image of the Drinfeld associator and thus the generating series of svMZVs:
\begin{equation}
	\Phi^{\sv}=\sv(\Phi)=\sum_{w\in\{e_0,e_1\}^\times}(-1)^{\# w}w\,\zeta^{\sv}_w=1-2\zeta(3)[e_0+e_1,[e_0,e_1]]+\ldots\ .
\end{equation}
%

%%%%%%%%%%%%%%%%%%%%%%%%%%%%%%%%%%%%%%%%%%%%%%%%%%%%%%%%%%%%%%%%%%%%%
%%%%%%%%%%%%%%%%%%%%%%%%%%%%%%%%%%%%%%%%%%%%%%%%%%%%%%%%%%%%%%%%%%%%%

\section{Explicit results for all weight three insertions}
\label{app:explicit3}
In this appendix we list the explicit results for all possible weight three polylogarithm insertions for the open-string setting, i.e.~the recursive relations for integrals $J_w^{s,t}$ for $w$ of weight three using the Drinfeld associator. For each case, we give the matrices $e_0$ and $e_1$ and the boundary value vectors $C_0$ and $C_1$, which are related via the Drinfeld recursion~\eqref{eqn:Drinrec}. Then we calculate $J_w^{s,t}$ using the shift relation \eqref{eqn:shiftrel} from the $J_w^{s,t+1}$ as discussed around \eqn{eqn:J001}. Deligne relations for the integrals $I_w^{s,t}$ are then given through the corresponding relation for $J_w^{s,t}$ upon replacing $\Phi$ by $\Phi^{\sv}$.

\paragraph{A.1. \boldmath$w\,{=}\,000$}
{\scriptsize
\begin{equation}
e_0\,{=}\,\!\left(\!
\begin{array}{cccccccc}
 s & s & 1 & 0 & 0 & 0 & 0 & 0 \\
 0 & 0 & 0 & 0 & 0 & 0 & 0 & 0 \\
 0 & 0 & s & s & 1 & 0 & 0 & 0 \\
 0 & 0 & 0 & 0 & 0 & 0 & 0 & 0 \\
 0 & 0 & 0 & 0 & s & s & 1 & 0 \\
 0 & 0 & 0 & 0 & 0 & 0 & 0 & 0 \\
 0 & 0 & 0 & 0 & 0 & 0 & s & s \\
 0 & 0 & 0 & 0 & 0 & 0 & 0 & 0 \\
\end{array}
\!\right)\!,\ 
e_1\,{=}\,\!\left(\!
\begin{array}{cccccccc}
 0 & 0 & 0 & 0 & 0 & 0 & 0 & 0 \\
 t & t & \frac{t}{s} & 0 & 0 & 0 & 0 & 0 \\
 0 & 0 & 0 & 0 & 0 & 0 & 0 & 0 \\
 0 & 0 & t & t & \frac{t}{s} & 0 & 0 & 0 \\
 0 & 0 & 0 & 0 & 0 & 0 & 0 & 0 \\
 0 & 0 & 0 & 0 & t & t & \frac{t}{s} & 0 \\
 0 & 0 & 0 & 0 & 0 & 0 & 0 & 0 \\
 0 & 0 & 0 & 0 & 0 & 0 & t & t \\
\end{array}
\!\right)\!,\ C_0\,{=}\,\!\left(\!\begin{array}{c}-\frac{1}{s^3}\\0\\\frac{1}{s^2}\\0\\-\frac{1}{s}\\0\\1\\0\end{array}\!\right)\!,\ 
C_1\,{=}\,\!\left(\!\begin{array}{c}s\,J_{000}^{s,t+1}\\{-}s\,J_{000}^{s,t+1}{-}J_{00}^{s,t+1}\\s\,J_{00}^{s,t+1}\\{-}s\,J_{00}^{s,t+1}{-}J_{0}^{s,t+1}\\s\,J_0^{s,t+1}\\{-}s\,J_{0}^{s,t+1}{-}J_e^{s,t+1}\\\frac{\Gamma(s+1)\Gamma(1+t)}{\Gamma(1+s+t)}\\1{-}\frac{\Gamma(s+1)\Gamma(1+t)}{\Gamma(1+s+t)}\end{array}\!\right)\!.
\end{equation}
}
\vspace{-2ex}
\begin{align}
	\implies J_{000}^{s,t}&=\frac{1}{t}\left((s+t)J_{000}^{s,t+1}+J_{00}^{s,t+1}\right)\\
	&=\frac{1}{st}(s+t,0,1,0,0,0,0,0)\cdot\Phi(e_0,e_1)\cdot \left(-\frac{1}{s^3},0,\frac{1}{s^2},0,-\frac{1}{s},0,1,0\right)^T.\notag
\end{align}

\paragraph{A.2 \boldmath$w\,{=}\,001$}
{\scriptsize
\begin{equation}
	e_0\,{=}\,\!\left(\!
	\begin{array}{cccccccc}
		s&s&1&0&0&0&0&0\\
		0&0&0&0&0&0&0&0\\
		0&0&s&s&1&0&0&0\\
		0&0&0&0&0&0&0&0\\
		0&0&0&0&s&s&0&\frac{s}{t}\\
		0&0&0&0&0&0&0&0\\
		0&0&0&0&0&0&s&s\\
		0&0&0&0&0&0&0&0
	\end{array}
	\!\right)\!,\ 
	e_1\,{=}\,\!\left(\!\begin{array}{cccccccc}
		0&0&0&0&0&0&0&0\\
		t&t&\frac{t}{s}&0&0&0&0&0\\
		0&0&0&0&0&0&0&0\\
		0&0&t&t&\frac{t}{s}&0&0&0\\
		0&0&0&0&0&0&0&0\\
		0&0&0&0&t&t&0&1\\
		0&0&0&0&0&0&0&0\\
		0&0&0&0&0&0&t&t
	\end{array}
	\!\right)\!,
	\ C_0\,{=}\,\!\left(\!\begin{array}{c}0\\0\\0\\0\\0\\0\\1\\0\end{array}\!\right)\!,\ 
	C_1\,{=}\,\!\left(\!\begin{array}{c}s\,J_{001}^{s,t+1}\\{-}s\,J_{001}^{s,t+1}{-}J_{01}^{s,t+1}{-}\zeta(3)\\s\,J_{01}^{s,t+1}\\{-}s\,J_{01}^{s,t+1}{-}J_{1}^{s,t+1}{-}\zeta(2)\\s\,J_{1}^{s,t+1}\\{-}s\,J_1^{s,t+1}{+}\frac{s}{t}J_e^{s,t+1}{-}\frac{1}{t}\\\frac{\Gamma(s+1)\Gamma(1+t)}{\Gamma(1+s+t)}\\1{-}\frac{\Gamma(s+1)\Gamma(1+t)}{\Gamma(1+s+t)}\end{array}\!\right)\!.
\end{equation}
}
\vspace{-2ex}
\begin{align}
	\implies J_{001}^{s,t}&=\frac{1}{t}\left((s+t)J_{001}^{s,t+1}+J_{01}^{s,t+1}\right)\\
	&=\frac{1}{st}(s+t,0,1,0,0,0,0,0)\cdot\Phi(e_0,e_1)\cdot (0,0,0,0,0,0,1,0)^T.\notag
\end{align}

\paragraph{A.3 \boldmath$w\,{=}\,010$}
{\scriptsize
\begin{equation}
e_0\,{=}\,\!\left(\!
\begin{array}{cccccccc}
	s & s & 1 & 0 & 0 & 0 & 0 & 0 \\
	0 & 0 & 0 & 0 & 0 & 0 & 0 & 0 \\
	0 & 0 & s & s & 0 & \frac{s}{t} & 0 & 0 \\
	0 & 0 & 0 & 0 & 0 & 0 & 0 & 0 \\
	0 & 0 & 0 & 0 & s & s & 1 & 0 \\
	0 & 0 & 0 & 0 & 0 & 0 & 0 & 0 \\
	0 & 0 & 0 & 0 & 0 & 0 & s & s \\
	0 & 0 & 0 & 0 & 0 & 0 & 0 & 0
\end{array}
\!\right)\!,\ 
e_1\,{=}\,\!\left(\!
\begin{array}{cccccccc}
	0 & 0 & 0 & 0 & 0 & 0 & 0 & 0 \\
	t & t & \frac{t}{s} & 0 & 0 & 0 & 0 & 0 \\
	0 & 0 & 0 & 0 & 0 & 0 & 0 & 0 \\
	0 & 0 & t & t & 0 & 1 & 0 & 0 \\
	0 & 0 & 0 & 0 & 0 & 0 & 0 & 0 \\
	0 & 0 & 0 & 0 & t & t & \frac{t}{s} & 0 \\
	0 & 0 & 0 & 0 & 0 & 0 & 0 & 0 \\
	0 & 0 & 0 & 0 & 0 & 0 & t & t \\
\end{array}
\!\right)\!,\ C_0\,{=}\,\!\left(\!\begin{array}{c}0\\0\\0\\0\\-\frac{1}{s}\\0\\1\\0\end{array}\!\right)\!,\ 
C_1\,{=}\,\!\left(\!\begin{array}{c}s\,J_{010}^{s,t+1}\\{-}s\,J_{010}^{s,t+1}{-}J_{10}^{s,t+1}{-}\zeta_{010}\\s\,J_{10}^{s,t+1}\\{-}s\,J_{01}^{s,t+1}{+}\frac{s}{t}J_0^{s,t+1}{+}\frac{1}{t}J_e^{s,t+1}{-}\zeta_{10}\\s\,J_{0}^{s,t+1}\\{-}s\,J_0^{s,t+1}{-}J_e^{s,t+1}\\\frac{\Gamma(s+1)\Gamma(1+t)}{\Gamma(1+s+t)}\\1{-}\frac{\Gamma(s+1)\Gamma(1+t)}{\Gamma(1+s+t)}\end{array}\!\right)\!,
\end{equation}
}
where $\zeta_{010}=-2\zeta(3)$ and $\zeta_{10}=-\zeta(2)$.
\begin{align}
	\implies J_{010}^{s,t}&=\frac{1}{t}\left((s+t)J_{010}^{s,t+1}+J_{10}^{s,t+1}\right)\\
	&=\frac{1}{st}(s+t,0,1,0,0,0,0,0)\cdot\Phi(e_0,e_1)\cdot \left(0,0,0,0,-\frac{1}{s},0,1,0\right)^T.\notag
\end{align}

\paragraph{A.4 \boldmath$w\,{=}\,100$}
{\scriptsize
\begin{equation}
e_0\,{=}\,\!\left(\!
\begin{array}{cccccccc}
 s & s & 0 & \frac{s}{t} & 0 & 0 & 0 & 0 \\
 0 & 0 & 0 & 0 & 0 & 0 & 0 & 0 \\
 0 & 0 & s & s & 1 & 0 & 0 & 0 \\
 0 & 0 & 0 & 0 & 0 & 0 & 0 & 0 \\
 0 & 0 & 0 & 0 & s & s & 1 & 0 \\
 0 & 0 & 0 & 0 & 0 & 0 & 0 & 0 \\
 0 & 0 & 0 & 0 & 0 & 0 & s & s \\
 0 & 0 & 0 & 0 & 0 & 0 & 0 & 0 \\
\end{array}
\!\right)\!,\ 
e_1\,{=}\,\!\left(\!
\begin{array}{cccccccc}
 0 & 0 & 0 & 0 & 0 & 0 & 0 & 0 \\
 t & t & 0 & 1 & 0 & 0 & 0 & 0 \\
 0 & 0 & 0 & 0 & 0 & 0 & 0 & 0 \\
 0 & 0 & t & t & \frac{t}{s} & 0 & 0 & 0 \\
 0 & 0 & 0 & 0 & 0 & 0 & 0 & 0 \\
 0 & 0 & 0 & 0 & t & t & \frac{t}{s} & 0 \\
 0 & 0 & 0 & 0 & 0 & 0 & 0 & 0 \\
 0 & 0 & 0 & 0 & 0 & 0 & t & t \\
\end{array}
\!\right)\!,\ C_0\,{=}\,\!\left(\!\begin{array}{c}0\\0\\\frac{1}{s^2}\\0\\-\frac{1}{s}\\0\\1\\0\end{array}\!\right)\!,\ 
C_1\,{=}\,\!\left(\!\begin{array}{c}s\,J_{100}^{s,t+1}\\{-}s\,J_{100}^{s,t+1}{+}\frac{s}{t}J_{00}^{s,t+1}-\zeta_{100}\\s\,J_{00}^{s,t+1}\\{-}s\,J_{00}^{s,t+1}{-}J_{0}^{s,t+1}\\s\,J_0^{s,t+1}\\{-}s\,J_0^{s,t+1}{-}J_e^{s,t+1} \\\frac{\Gamma(s+1)\Gamma(1+t)}{\Gamma(1+s+t)}\\1{-}\frac{\Gamma(s+1)\Gamma(1+t)}{\Gamma(1+s+t)}\\\end{array}\!\right)\!,
\end{equation}
}
where $\zeta_{100}=\zeta(3)$.
\begin{align}
	\implies J_{100}^{s,t}&=\frac{1}{t}\left((s+t)J_{100}^{s,t+1}-\frac{s}{t}J_{00}^{s,t+1}-\frac{1}{t}J_0^{s,t+1}\right)\\
	&=\frac{1}{st}\left(s+t,0,-\frac{s}{t},0,-\frac{1}{t},0,0,0\right)\cdot\Phi(e_0,e_1)\cdot \left(0,0,\frac{1}{s^2},0,-\frac{1}{s},0,1,0\right)^T.\notag
\end{align}
\paragraph{A.5 \boldmath$w\,{=}\,011$}
{\scriptsize
\begin{equation}
e_0\,{=}\,\!\left(\!
\begin{array}{cccccccc}
 s & s & 1 & 0 & 0 & 0 & 0 & 0 \\
 0 & 0 & 0 & 0 & 0 & 0 & 0 & 0 \\
 0 & 0 & s & s & 0 & \frac{s}{t} & 0 & 0 \\
 0 & 0 & 0 & 0 & 0 & 0 & 0 & 0 \\
 0 & 0 & 0 & 0 & s & s & 0 & \frac{s}{t} \\
 0 & 0 & 0 & 0 & 0 & 0 & 0 & 0 \\
 0 & 0 & 0 & 0 & 0 & 0 & s & s \\
 0 & 0 & 0 & 0 & 0 & 0 & 0 & 0 \\
\end{array}
\!\right)\!,\ 
e_1\,{=}\,\!\left(\!
\begin{array}{cccccccc}
 0 & 0 & 0 & 0 & 0 & 0 & 0 & 0 \\
 t & t & \frac{t}{s} & 0 & 0 & 0 & 0 & 0 \\
 0 & 0 & 0 & 0 & 0 & 0 & 0 & 0 \\
 0 & 0 & t & t & 0 & 1 & 0 & 0 \\
 0 & 0 & 0 & 0 & 0 & 0 & 0 & 0 \\
 0 & 0 & 0 & 0 & t & t & 0 & 1 \\
 0 & 0 & 0 & 0 & 0 & 0 & 0 & 0 \\
 0 & 0 & 0 & 0 & 0 & 0 & t & t \\
\end{array}
\!\right)\!,\ C_0\,{=}\,\!\left(\!\begin{array}{c}0\\0\\0\\0\\0\\0\\1\\0\end{array}\!\right)\!,\ 
C_1\,{=}\,\!\left(\!\begin{array}{c}s\,J_{011}^{s,t+1}\\{-}s\,J_{011}^{s,t+1}{-}J_{11}^{s,t+1}{+}\zeta_{011}\\s\,J_{11}^{s,t+1}\\{-}s\,J_{11}^{s,t+1}{+}\frac{s}{t}J_1^{s,t+1}{-}\frac{s}{t^2}J_e^{s,t+1}{+}\frac{1}{t^2}\\s\,J_{1}^{s,t+1}\\{-}s\,J_1^{s,t+1}{+}\frac{s}{t}J_e^{s,t+1}{-}\frac{1}{t}\\\frac{\Gamma(s+1)\Gamma(1+t)}{\Gamma(1+s+t)}\\1{-}\frac{\Gamma(s+1)\Gamma(1+t)}{\Gamma(1+s+t)}\end{array}\!\right)\!,
\end{equation}
}
where $\zeta_{011}=\zeta(3)$.
\begin{align}
	\implies J_{011}^{s,t}&=\frac{1}{t}\left((s+t)J_{011}^{s,t+1}+J_{11}^{s,t+1}\right)\\
	&=\frac{1}{st}(s+t,0,1,0,0,0,0,0)\cdot\Phi(e_0,e_1)\cdot (0,0,0,0,0,0,1,0)^T.\notag
\end{align}

\paragraph{A.6 \boldmath$w\,{=}\,101$}
{\scriptsize
\begin{equation}
e_0\,{=}\,\!\left(\!
\begin{array}{cccccccc}
 s & s & 0 & \frac{s}{t} & 0 & 0 & 0 & 0 \\
 0 & 0 & 0 & 0 & 0 & 0 & 0 & 0 \\
 0 & 0 & s & s & 1 & 0 & 0 & 0 \\
 0 & 0 & 0 & 0 & 0 & 0 & 0 & 0 \\
 0 & 0 & 0 & 0 & s & s & 0 & \frac{s}{t} \\
 0 & 0 & 0 & 0 & 0 & 0 & 0 & 0 \\
 0 & 0 & 0 & 0 & 0 & 0 & s & s \\
 0 & 0 & 0 & 0 & 0 & 0 & 0 & 0 \\
\end{array}
\!\right)\!,\ 
e_1\,{=}\,\!\left(\!
\begin{array}{cccccccc}
 0 & 0 & 0 & 0 & 0 & 0 & 0 & 0 \\
 t & t & 0 & 1 & 0 & 0 & 0 & 0 \\
 0 & 0 & 0 & 0 & 0 & 0 & 0 & 0 \\
 0 & 0 & t & t & \frac{t}{s} & 0 & 0 & 0 \\
 0 & 0 & 0 & 0 & 0 & 0 & 0 & 0 \\
 0 & 0 & 0 & 0 & t & t & 0 & 1 \\
 0 & 0 & 0 & 0 & 0 & 0 & 0 & 0 \\
 0 & 0 & 0 & 0 & 0 & 0 & t & t \\
\end{array}
\!\right)\!,\ C_0\,{=}\,\!\left(\!\begin{array}{c}0\\0\\0\\0\\0\\0\\1\\0\end{array}\!\right)\!,\ 
C_1\,{=}\,\!\left(\!\begin{array}{c}s\,J_{101}^{s,t+1}\\{-}s\,J_{101}^{s,t+1}{-}\frac{s}{t}J_{01}^{s,t+1}{+}\frac{1}{t}J_1^{s,t+1}{+}\frac{\zeta(2)}{t}{+}\zeta_{101}\\s\,J_{01}^{s,t+1}\\{-}s\,J_{01}^{s,t+1}{-}s\,J_1^{s,t+1}{-}\zeta(2)\\s\,J_{1}^{s,t+1}\\{-}s\,J_1^{s,t+1}{+}\frac{s}{t}J_e^{s,t+1}-\frac{1}{t}\\\frac{\Gamma(s+1)\Gamma(1+t)}{\Gamma(1+s+t)}\\1{-}\frac{\Gamma(s+1)\Gamma(1+t)}{\Gamma(1+s+t)}\end{array}\!\right)\!,
\end{equation}
}
where $\zeta_{101}=-2\zeta(3)$.
\begin{align}
	\implies J_{101}^{s,t}&=\frac{1}{t}\left((s+t)J_{101}^{s,t+1}-\frac{s}{t}J_{01}^{s,t+1}-\frac{1}{t}J_1^{s,t+1}\right)\\
	&=\frac{1}{st}\left(s+t,0,-\frac{s}{t},0,-\frac{1}{t},0,0,0\right)\cdot\Phi(e_0,e_1)\cdot (0,0,0,0,0,0,1,0)^T.\notag
\end{align}

\paragraph{A.7 \boldmath$w\,{=}\,110$}
{\scriptsize
\begin{equation}
e_0\,{=}\,\!\left(\!
\begin{array}{cccccccc}
 s & s & 0 & \frac{s}{t} & 0 & 0 & 0 & 0 \\
 0 & 0 & 0 & 0 & 0 & 0 & 0 & 0 \\
 0 & 0 & s & s & 0 & \frac{s}{t} & 0 & 0 \\
 0 & 0 & 0 & 0 & 0 & 0 & 0 & 0 \\
 0 & 0 & 0 & 0 & s & s & 1 & 0 \\
 0 & 0 & 0 & 0 & 0 & 0 & 0 & 0 \\
 0 & 0 & 0 & 0 & 0 & 0 & s & s \\
 0 & 0 & 0 & 0 & 0 & 0 & 0 & 0 \\
\end{array}
\!\right)\!,\ 
e_1\,{=}\,\!\left(\!
\begin{array}{cccccccc}
 0 & 0 & 0 & 0 & 0 & 0 & 0 & 0 \\
 t & t & 0 & 1 & 0 & 0 & 0 & 0 \\
 0 & 0 & 0 & 0 & 0 & 0 & 0 & 0 \\
 0 & 0 & t & t & 0 & 1 & 0 & 0 \\
 0 & 0 & 0 & 0 & 0 & 0 & 0 & 0 \\
 0 & 0 & 0 & 0 & t & t & \frac{t}{s} & 0 \\
 0 & 0 & 0 & 0 & 0 & 0 & 0 & 0 \\
 0 & 0 & 0 & 0 & 0 & 0 & t & t \\
\end{array}
\!\right)\!,\ C_0\,{=}\,\!\left(\!\begin{array}{c}0\\0\\0\\0\\-\frac{1}{s}\\0\\1\\0\end{array}\!\right)\!,\ 
C_1\,{=}\,\!\left(\!\!\!\begin{array}{c}s\,J_{110}^{s,t+1}\\{-}s\,J_{110}^{s,t+1}{+}\frac{s}{t}J_{10}^{s,t+1}{-}\frac{s}{t^2}J_0^{s,t+1}{-}\frac{1}{t^2}J_e^{s,t+1}{-}\frac{\zeta(2)}{t}{+}\zeta_{110}\\s\,J_{10}^{s,t+1}\\{-}s\,J_{10}^{s,t+1}{+}\frac{s}{t}J_0^{s,t+1}{+}\frac{1}{t}J_e^{s,t+1}{+}\zeta(2)\\s\,J_{0}^{s,t+1}\\{-}s\,J_0^{s,t+1}{-}J_e^{s,t+1}\\\frac{\Gamma(s+1)\Gamma(1+t)}{\Gamma(1+s+t)}\\1{-}\frac{\Gamma(s+1)\Gamma(1+t)}{\Gamma(1+s+t)}\end{array}\!\!\!\right)\!\!,
\end{equation}
}
where $\zeta_{110}=\zeta(3)$.
\begin{align}
	\implies J_{110}^{s,t}&=\frac{1}{t}\left((s+t)J_{110}^{s,t+1}-\frac{s}{t}J_{10}^{s,t+1}+\frac{s}{t^2}J_0^{s,t+1}+\frac{1}{t^2}J_e^{s,t+1}\right)\\
	&=\frac{1}{st}\left(s+t,0,-\frac{s}{t},0,\frac{s}{t^2},0,\frac{1}{t^2},0\right)\cdot\Phi(e_0,e_1)\cdot \left(0,0,0,0,-\frac{1}{s},0,1,0\right)^T.\notag
\end{align}

\paragraph{A.8 \boldmath$w\,{=}\,111$}
{\scriptsize
\begin{equation}
e_0\,{=}\,\!\left(\!
\begin{array}{cccccccc}
 s & s & 0 & \frac{s}{t} & 0 & 0 & 0 & 0 \\
 0 & 0 & 0 & 0 & 0 & 0 & 0 & 0 \\
 0 & 0 & s & s & 0 & \frac{s}{t} & 0 & 0 \\
 0 & 0 & 0 & 0 & 0 & 0 & 0 & 0 \\
 0 & 0 & 0 & 0 & s & s & 0 & \frac{s}{t} \\
 0 & 0 & 0 & 0 & 0 & 0 & 0 & 0 \\
 0 & 0 & 0 & 0 & 0 & 0 & s & s \\
 0 & 0 & 0 & 0 & 0 & 0 & 0 & 0 \\
\end{array}
\!\right)\!,\ 
e_1\,{=}\,\!\left(\!
\begin{array}{cccccccc}
 0 & 0 & 0 & 0 & 0 & 0 & 0 & 0 \\
 t & t & 0 & 1 & 0 & 0 & 0 & 0 \\
 0 & 0 & 0 & 0 & 0 & 0 & 0 & 0 \\
 0 & 0 & t & t & 0 & 1 & 0 & 0 \\
 0 & 0 & 0 & 0 & 0 & 0 & 0 & 0 \\
 0 & 0 & 0 & 0 & t & t & 0 & 1 \\
 0 & 0 & 0 & 0 & 0 & 0 & 0 & 0 \\
 0 & 0 & 0 & 0 & 0 & 0 & t & t \\
 \end{array}
\!\right)\!,\ C_0\,{=}\,\!\left(\!\begin{array}{c}0\\0\\0\\0\\0\\0\\1\\0\end{array}\!\right)\!,\ 
C_1\,{=}\,\!\left(\!\begin{array}{c}s\,J_{111}^{s,t+1}\\{-}s\,J_{111}^{s,t+1}{+}\frac{s}{t}J_{11}^{s,t+1}{-}\frac{s}{t^2}J_1^{s,t+1}{+}\frac{s}{t^3}J_e^{s,t+1}-\frac{1}{t^3}\\s\,J_{11}^{s,t+1}\\{-}s\,J_{11}^{s,t+1}{+}\frac{s}{t}J_1^{s,t+1}{-}\frac{s}{t^2}J_e^{s,t+1}{+}\frac{1}{t^2}\\s\,J_{1}^{s,t+1}\\{-}s\,J_1^{s,t+1}{+}\frac{s}{t}J_e^{s,t+1}{-}\frac{1}{t}\\\frac{\Gamma(s+1)\Gamma(1+t)}{\Gamma(1+s+t)}\\1{-}\frac{\Gamma(s+1)\Gamma(1+t)}{\Gamma(1+s+t)}\end{array}\!\right)\!.
\end{equation}
}
\begin{align}
	\implies J_{111}^{s,t}&=\frac{1}{t}\left((s+t)J_{111}^{s,t+1}-\frac{s}{t}J_{00}^{s,t+1}+\frac{s}{t^2}J_0^{s,t+1}-\frac{s}{t^3}J_e^{s,t+1}\right)\\
	&=\frac{1}{st}\left(s+t,0,-\frac{s}{t},0,\frac{s}{t^2},0,-\frac{s}{t^3},0\right)\cdot\Phi(e_0,e_1)\cdot (0,0,0,0,0,0,1,0)^T.\notag
\end{align}

\section{Matrices for five-point example}
\label{app:5pt}
The matrices from \exref{ex:3} are (after setting $r_3\,{=}\,r_4\,{=}\,0$)
{\tiny
\begin{subequations}
\begin{align}
	e_0&=\left(
\begin{array}{cccccccccccc}
 s_{03}{+}s_{04}{+}s_{34} & s_{04} & 0 & s_{03} & 0 & 0 & 0 & 0 & 0 & \frac{s_{03}}{s_{13}} & 0 & 0 \\
 0 & s_{03} & 0 & 0 & s_{03} \left(\frac{s_{34}}{s_{13}}{+}1\right) & {-}\frac{s_{03} s_{14}}{s_{13}} & 0 & 0 & 0 & 0 & \frac{s_{03}}{s_{13}} & 0 \\
 {-}s_{03}{-}s_{34} & {-}s_{03} & s_{04} & {-}s_{03} & {-}\frac{s_{03} \left(s_{13}{+}s_{34}\right)}{s_{13}} & \frac{s_{03} s_{14}}{s_{13}} & 0 & 0 & 0 & {-}\frac{s_{03}}{s_{13}} & {-}\frac{s_{03}}{s_{13}} & 0 \\
 0 & 0 & 0 & s_{04} & s_{04} & s_{04} & 0 & 0 & 0 & 0 & 0 & 0 \\
 0 & 0 & 0 & 0 & 0 & 0 & 0 & 0 & 0 & 0 & 0 & 0 \\
 0 & 0 & 0 & 0 & 0 & 0 & 0 & 0 & 0 & 0 & 0 & 0 \\
 0 & 0 & 0 & 0 & 0 & 0 & s_{03}{+}s_{04}{+}s_{34} & s_{04} & 0 & s_{03} & 0 & 0 \\
 0 & 0 & 0 & 0 & 0 & 0 & 0 & s_{03} & 0 & 0 & s_{03} \left(\frac{s_{34}}{s_{13}}{+}1\right) & {-}\frac{s_{03} s_{14}}{s_{13}} \\
 0 & 0 & 0 & 0 & 0 & 0 & {-}s_{03}{-}s_{34} & {-}s_{03} & s_{04} & {-}s_{03} & {-}\frac{s_{03} \left(s_{13}{+}s_{34}\right)}{s_{13}} & \frac{s_{03} s_{14}}{s_{13}} \\
 0 & 0 & 0 & 0 & 0 & 0 & 0 & 0 & 0 & s_{04} & s_{04} & s_{04} \\
 0 & 0 & 0 & 0 & 0 & 0 & 0 & 0 & 0 & 0 & 0 & 0 \\
 0 & 0 & 0 & 0 & 0 & 0 & 0 & 0 & 0 & 0 & 0 & 0
\end{array}
\right),\\
e_1&=\left(
\begin{array}{cccccccccccc}
 0 & 0 & 0 & 0 & 0 & 0 & 0 & 0 & 0 & 0 & 0 & 0 \\
 s_{14} & s_{14} & s_{14} & 0 & 0 & 0 & 0 & 0 & 0 & 0 & 0 & 0 \\
 0 & 0 & 0 & 0 & 0 & 0 & 0 & 0 & 0 & 0 & 0 & 0 \\
 s_{13} \left(\frac{s_{34}}{s_{03}}{+}1\right) & 0 & {-}\frac{s_{04} s_{13}}{s_{03}} & s_{13} & 0 & 0 & 0 & 0 & 0 & 1 & 0 & 0 \\
 0 & s_{13} & 0 & s_{14} & s_{13}+s_{14}+s_{34} & 0 & 0 & 0 & 0 & 0 & 1 & 0 \\
 {-}\frac{s_{13} s_{34}}{s_{03}} & 0 & \left(\frac{s_{04}}{s_{03}}{+}1\right) s_{13} & s_{34} & 0 & s_{13}{+}s_{14}{+}s_{34} & 0 & 0 & 0 & 0 & 0 & 1 \\
 0 & 0 & 0 & 0 & 0 & 0 & 0 & 0 & 0 & 0 & 0 & 0 \\
 0 & 0 & 0 & 0 & 0 & 0 & s_{14} & s_{14} & s_{14} & 0 & 0 & 0 \\
 0 & 0 & 0 & 0 & 0 & 0 & 0 & 0 & 0 & 0 & 0 & 0 \\
 0 & 0 & 0 & 0 & 0 & 0 & s_{13} \left(\frac{s_{34}}{s_{03}}{+}1\right) & 0 & {-}\frac{s_{04} s_{13}}{s_{03}} & s_{13} & 0 & 0 \\
 0 & 0 & 0 & 0 & 0 & 0 & 0 & s_{13} & 0 & s_{14} & s_{13}{+}s_{14}{+}s_{34} & 0 \\
 0 & 0 & 0 & 0 & 0 & 0 & {-}\frac{s_{13} s_{34}}{s_{03}} & 0 & \left(\frac{s_{04}}{s_{03}}{+}1\right) s_{13} & s_{34} & 0 & s_{13}{+}s_{14}{+}s_{34}
\end{array}
\right).
\end{align}
\end{subequations}
}
%

% Bibliography

%% [A] Recommended: using JHEP.bst file
\bibliographystyle{JHEP}
\bibliography{biblio.bib}

\def\cprime{$'$}

\providecommand{\href}[2]{#2}\begingroup\raggedright\begin{thebibliography}{10}

\bibitem{Veneziano:1968yb}
G.~Veneziano, \emph{{Construction of a crossing-symmetric, Regge behaved
  amplitude for linearly rising trajectories}},
  \href{https://doi.org/10.1007/BF02824451}{\emph{Nuovo Cim. A} {\bfseries 57}
  (1968) 190}.

\bibitem{Virasoro:1969me}
M.A.~Virasoro, \emph{{Alternative constructions of crossing-symmetric
  amplitudes with Regge behavior}},
  \href{https://doi.org/10.1103/PhysRev.177.2309}{\emph{Phys. Rev.} {\bfseries
  177} (1969) 2309}.

\bibitem{Shapiro:1970gy}
J.A.~Shapiro, \emph{{Electrostatic analog for the Virasoro model}},
  \href{https://doi.org/10.1016/0370-2693(70)90255-8}{\emph{Phys. Lett. B}
  {\bfseries 33} (1970) 361}.

\bibitem{Kawai:1985xq}
H.~Kawai, D.C.~Lewellen and S.H.H.~Tye, \emph{{A Relation Between Tree
  Amplitudes of Closed and Open Strings}},
  \href{https://doi.org/10.1016/0550-3213(86)90362-7}{\emph{Nucl. Phys.}
  {\bfseries B269} (1986) 1}.

\bibitem{Brown:2019wna}
F.~Brown and C.~Dupont, \emph{{Single-valued integration and superstring
  amplitudes in genus zero}},
  \href{https://doi.org/10.1007/s00220-021-03969-4}{\emph{Commun. Math. Phys.}
  {\bfseries 382} (2021) 815}
  [\href{https://arxiv.org/abs/1910.01107}{{\ttfamily 1910.01107}}].

\bibitem{Mizera:2017cqs}
S.~Mizera, \emph{{Combinatorics and Topology of Kawai-Lewellen-Tye Relations}},
  \href{https://doi.org/10.1007/JHEP08(2017)097}{\emph{JHEP} {\bfseries 08}
  (2017) 097} [\href{https://arxiv.org/abs/1706.08527}{{\ttfamily
  1706.08527}}].

\bibitem{Mizera:2019gea}
S.~Mizera, \emph{{Aspects of Scattering Amplitudes and Moduli Space
  Localization}}, Ph.D. thesis, Princeton, Inst. Advanced Study, 2020.
\newblock \href{https://arxiv.org/abs/1906.02099}{{\ttfamily 1906.02099}}.
\newblock 10.1007/978-3-030-53010-5.

\bibitem{Stieberger:2022lss}
S.~Stieberger, \emph{{A Relation between One-Loop Amplitudes of Closed and Open
  Strings (One-Loop KLT Relation)}},
  \href{https://arxiv.org/abs/2212.06816}{{\ttfamily 2212.06816}}.

\bibitem{Stieberger:2023nol}
S.~Stieberger, \emph{{One-Loop Double Copy Relation in String Theory}},
  \href{https://doi.org/10.1103/PhysRevLett.132.191602}{\emph{Phys. Rev. Lett.}
  {\bfseries 132} (2024) 191602}
  [\href{https://arxiv.org/abs/2310.07755}{{\ttfamily 2310.07755}}].

\bibitem{Mazloumi:2024wys}
P.~Mazloumi and S.~Stieberger, \emph{{One-loop double copy relation from
  twisted (co)homology}},
  \href{https://doi.org/10.1007/JHEP10(2024)148}{\emph{JHEP} {\bfseries 10}
  (2024) 148} [\href{https://arxiv.org/abs/2403.05208}{{\ttfamily
  2403.05208}}].

\bibitem{Mafra:2018qqe}
C.R.~Mafra and O.~Schlotterer, \emph{{Towards the n-point one-loop superstring
  amplitude. Part III. One-loop correlators and their double-copy structure}},
  \href{https://doi.org/10.1007/JHEP08(2019)092}{\emph{JHEP} {\bfseries 08}
  (2019) 092} [\href{https://arxiv.org/abs/1812.10971}{{\ttfamily
  1812.10971}}].

\bibitem{Stieberger:2013wea}
S.~Stieberger, \emph{{Closed superstring amplitudes, single-valued multiple
  zeta values and the Deligne associator}},
  \href{https://doi.org/10.1088/1751-8113/47/15/155401}{\emph{J.Phys.}
  {\bfseries A47} (2014) 155401}
  [\href{https://arxiv.org/abs/1310.3259}{{\ttfamily 1310.3259}}].

\bibitem{BrownSVMPL}
F.~Brown, \emph{Single-valued multiple polylogarithms in one variable},
  {\emph{C.R. Acad. Sci. Paris} {\bfseries 338} (2004) 527}.

\bibitem{Brown:2013gia}
F.~Brown, \emph{{Single-valued Motivic Periods and Multiple Zeta Values}},
  \href{https://doi.org/10.1017/fms.2014.18}{\emph{SIGMA} {\bfseries 2} (2014)
  e25} [\href{https://arxiv.org/abs/1309.5309}{{\ttfamily 1309.5309}}].

\bibitem{Stieberger:2014hba}
S.~Stieberger and T.R.~Taylor, \emph{{Closed String Amplitudes as Single-Valued
  Open String Amplitudes}},
  \href{https://doi.org/10.1016/j.nuclphysb.2014.02.005}{\emph{Nucl.Phys.}
  {\bfseries B881} (2014) 269}
  [\href{https://arxiv.org/abs/1401.1218}{{\ttfamily 1401.1218}}].

\bibitem{Brown:2018omk}
F.~Brown and C.~Dupont, \emph{{Single-valued integration and double copy}},
  \href{https://doi.org/10.1515/crelle-2020-0042}{\emph{J. Reine Angew. Math.}
  {\bfseries 2021} (2021) 145}
  [\href{https://arxiv.org/abs/1810.07682}{{\ttfamily 1810.07682}}].

\bibitem{Broedel:2013aza}
J.~Broedel, O.~Schlotterer, S.~Stieberger and T.~Terasoma, \emph{{All order
  $\alpha'$-expansion of superstring trees from the Drinfeld associator}},
  \href{https://doi.org/10.1103/PhysRevD.89.066014}{\emph{Phys.Rev.} {\bfseries
  D89} (2014) 066014} [\href{https://arxiv.org/abs/1304.7304}{{\ttfamily
  1304.7304}}].

\bibitem{BK}
J.~Broedel and A.~Kaderli, \emph{{Amplitude recursions with an extra marked
  point}}, \href{https://doi.org/10.4310/CNTP.2022.v16.n1.a3}{\emph{Commun.
  Num. Theor. Phys.} {\bfseries 16} (2022) 75}
  [\href{https://arxiv.org/abs/1912.09927}{{\ttfamily 1912.09927}}].

\bibitem{Kaderli}
A.~Kaderli, \emph{{A note on the Drinfeld associator for genus-zero superstring
  amplitudes in twisted de Rham theory}},
  \href{https://doi.org/10.1088/1751-8121/ab9462}{\emph{J. Phys. A} {\bfseries
  53} (2020) 415401} [\href{https://arxiv.org/abs/1912.09406}{{\ttfamily
  1912.09406}}].

\bibitem{Baune:2024uwj}
K.~Baune, J.~Broedel and F.~Zerbini, \emph{{Closed-string amplitude recursions
  from the Deligne associator}},
  \href{https://arxiv.org/abs/2412.17579}{{\ttfamily 2412.17579}}.

\bibitem{Alday:2022uxp}
L.F.~Alday, T.~Hansen and J.A.~Silva, \emph{{AdS Virasoro-Shapiro from
  dispersive sum rules}},
  \href{https://doi.org/10.1007/JHEP10(2022)036}{\emph{JHEP} {\bfseries 10}
  (2022) 036} [\href{https://arxiv.org/abs/2204.07542}{{\ttfamily
  2204.07542}}].

\bibitem{Alday:2022xwz}
L.F.~Alday, T.~Hansen and J.A.~Silva, \emph{{AdS Virasoro-Shapiro from
  single-valued periods}},
  \href{https://doi.org/10.1007/JHEP12(2022)010}{\emph{JHEP} {\bfseries 12}
  (2022) 010} [\href{https://arxiv.org/abs/2209.06223}{{\ttfamily
  2209.06223}}].

\bibitem{Alday:2023jdk}
L.F.~Alday, T.~Hansen and J.A.~Silva, \emph{{Emergent Worldsheet for the AdS
  Virasoro-Shapiro Amplitude}},
  \href{https://doi.org/10.1103/PhysRevLett.131.161603}{\emph{Phys. Rev. Lett.}
  {\bfseries 131} (2023) 161603}
  [\href{https://arxiv.org/abs/2305.03593}{{\ttfamily 2305.03593}}].

\bibitem{Alday:2023mvu}
L.F.~Alday and T.~Hansen, \emph{{The AdS Virasoro-Shapiro amplitude}},
  \href{https://doi.org/10.1007/JHEP10(2023)023}{\emph{JHEP} {\bfseries 10}
  (2023) 023} [\href{https://arxiv.org/abs/2306.12786}{{\ttfamily
  2306.12786}}].

\bibitem{Alday:2024xpq}
L.F.~Alday, M.~Nocchi, C.~Virally and X.~Zhou, \emph{{On the Regge behaviour of
  the AdS Virasoro-Shapiro amplitude}},
  \href{https://doi.org/10.1007/JHEP04(2025)064}{\emph{JHEP} {\bfseries 04}
  (2025) 064} [\href{https://arxiv.org/abs/2409.03695}{{\ttfamily
  2409.03695}}].

\bibitem{Alday:2024rjs}
L.F.~Alday, G.~Giribet and T.~Hansen, \emph{{On the AdS$_{3}$ Virasoro-Shapiro
  amplitude}}, \href{https://doi.org/10.1007/JHEP03(2025)002}{\emph{JHEP}
  {\bfseries 03} (2025) 002}
  [\href{https://arxiv.org/abs/2412.05246}{{\ttfamily 2412.05246}}].

\bibitem{Chester:2024esn}
S.M.~Chester, T.~Hansen and D.-l.~Zhong, \emph{{The type IIA Virasoro-Shapiro
  amplitude in AdS$_{4}$$\times$CP$^{3}$ from ABJM theory}},
  \href{https://doi.org/10.1007/JHEP05(2025)040}{\emph{JHEP} {\bfseries 05}
  (2025) 040} [\href{https://arxiv.org/abs/2412.08689}{{\ttfamily
  2412.08689}}].

\bibitem{Chester:2024wnb}
S.M.~Chester and D.-l.~Zhong, \emph{{AdS$_3\times$S$^3$ Virasoro-Shapiro
  Amplitude with Ramond-Ramond Flux}},
  \href{https://doi.org/10.1103/PhysRevLett.134.151602}{\emph{Phys. Rev. Lett.}
  {\bfseries 134} (2025) 151602}
  [\href{https://arxiv.org/abs/2412.06429}{{\ttfamily 2412.06429}}].

\bibitem{Wang:2025pjo}
B.~Wang, D.~Wu and E.Y.~Yuan, \emph{{The Kaluza-Klein AdS Virasoro-Shapiro
  Amplitude near Flat Space}},
  \href{https://arxiv.org/abs/2503.01964}{{\ttfamily 2503.01964}}.

\bibitem{Alday:2024ksp}
L.F.~Alday and T.~Hansen, \emph{{Single-valuedness of the AdS Veneziano
  amplitude}}, \href{https://doi.org/10.1007/JHEP08(2024)108}{\emph{JHEP}
  {\bfseries 08} (2024) 108}
  [\href{https://arxiv.org/abs/2404.16084}{{\ttfamily 2404.16084}}].

\bibitem{Alday:2024yax}
L.F.~Alday, S.M.~Chester, T.~Hansen and D.-l.~Zhong, \emph{{The AdS Veneziano
  amplitude at small curvature}},
  \href{https://doi.org/10.1007/JHEP05(2024)322}{\emph{JHEP} {\bfseries 05}
  (2024) 322} [\href{https://arxiv.org/abs/2403.13877}{{\ttfamily
  2403.13877}}].

\bibitem{Alday:2025bjp}
L.F.~Alday, M.~Nocchi and A.S.~Sangar\'e, \emph{{Stringy KLT Relations on
  $AdS$}},  \href{https://arxiv.org/abs/2504.19973}{{\ttfamily 2504.19973}}.

\bibitem{Brownhyperlogs}
F.~Brown, \emph{Single-valued hyperlogarithms and unipotent differential
  equations}, {\emph{Preprint (https://www.ihes.fr/~brown/RHpaper5.pdf)} (2004)
  }.

\bibitem{Terasoma}
T.~Terasoma, \emph{{Selberg Integrals and Multiple Zeta Values}},
  {\emph{Compositio Mathematica} {\bfseries 133} (2002) 1}.

\bibitem{VanZerb}
P.~Vanhove and F.~Zerbini, \emph{{Single-valued hyperlogarithms, correlation
  functions and closed string amplitudes}},
  \href{https://doi.org/10.4310/ATMP.2022.v26.n2.a5}{\emph{Adv. Theor. Math.
  Phys.} {\bfseries 26} (2022) 455}
  [\href{https://arxiv.org/abs/1812.03018}{{\ttfamily 1812.03018}}].

\bibitem{DelDuca:2016lad}
V.~Del~Duca, S.~Druc, J.~Drummond, C.~Duhr, F.~Dulat, R.~Marzucca et~al.,
  \emph{{Multi-Regge kinematics and the moduli space of Riemann spheres with
  marked points}}, \href{https://doi.org/10.1007/JHEP08(2016)152}{\emph{JHEP}
  {\bfseries 08} (2016) 152}
  [\href{https://arxiv.org/abs/1606.08807}{{\ttfamily 1606.08807}}].

\bibitem{Broedel:2016kls}
J.~Broedel, M.~Sprenger and A.~Torres~Orjuela, \emph{{Towards single-valued
  polylogarithms in two variables for the seven-point remainder function in
  multi-Regge kinematics}},
  \href{https://doi.org/10.1016/j.nuclphysb.2016.12.016}{\emph{Nucl. Phys. B}
  {\bfseries 915} (2017) 394}
  [\href{https://arxiv.org/abs/1606.08411}{{\ttfamily 1606.08411}}].

\bibitem{Frost:2023stm}
H.~Frost, M.~Hidding, D.~Kamlesh, C.~Rodriguez, O.~Schlotterer and B.~Verbeek,
  \emph{{Motivic coaction and single-valued map of polylogarithms from zeta
  generators}}, \href{https://doi.org/10.1088/1751-8121/ad5edf}{\emph{J. Phys.
  A} {\bfseries 57} (2024) 31LT01}
  [\href{https://arxiv.org/abs/2312.00697}{{\ttfamily 2312.00697}}].

\bibitem{Frost:2025lre}
H.~Frost, M.~Hidding, D.~Kamlesh, C.~Rodriguez, O.~Schlotterer and B.~Verbeek,
  \emph{{Deriving motivic coactions and single-valued maps at genus zero from
  zeta generators}},  \href{https://arxiv.org/abs/2503.02096}{{\ttfamily
  2503.02096}}.

\bibitem{Mizera:2016jhj}
S.~Mizera, \emph{{Inverse of the String Theory KLT Kernel}},
  \href{https://doi.org/10.1007/JHEP06(2017)084}{\emph{JHEP} {\bfseries 06}
  (2017) 084} [\href{https://arxiv.org/abs/1610.04230}{{\ttfamily
  1610.04230}}].

\bibitem{Mizera:2017rqa}
S.~Mizera, \emph{{Scattering Amplitudes from Intersection Theory}},
  \href{https://doi.org/10.1103/PhysRevLett.120.141602}{\emph{Phys. Rev. Lett.}
  {\bfseries 120} (2018) 141602}
  [\href{https://arxiv.org/abs/1711.00469}{{\ttfamily 1711.00469}}].

\bibitem{Mafra:2019ddf}
C.R.~Mafra and O.~Schlotterer, \emph{{All Order $\alpha'$ Expansion of One-Loop
  Open-String Integrals}},
  \href{https://doi.org/10.1103/PhysRevLett.124.101603}{\emph{Phys. Rev. Lett.}
  {\bfseries 124} (2020) 101603}
  [\href{https://arxiv.org/abs/1908.09848}{{\ttfamily 1908.09848}}].

\bibitem{Mafra:2019xms}
C.R.~Mafra and O.~Schlotterer, \emph{{One-loop open-string integrals from
  differential equations: all-order $\alpha'$-expansions at $n$ points}},
  \href{https://doi.org/10.1007/JHEP03(2020)007}{\emph{JHEP} {\bfseries 03}
  (2020) 007} [\href{https://arxiv.org/abs/1908.10830}{{\ttfamily
  1908.10830}}].

\bibitem{Broedel:2020tmd}
J.~Broedel, A.~Kaderli and O.~Schlotterer, \emph{{Two dialects for KZB
  equations: generating one-loop open-string integrals}},
  \href{https://doi.org/10.1007/JHEP12(2020)036}{\emph{JHEP} {\bfseries 12}
  (2020) 036}.

\bibitem{Brown:2013qva}
F.~Brown, \emph{{Iterated integrals in quantum field theory}},  in
  \emph{{Proceedings, Geometric and Topological Methods for Quantum Field
  Theory : 6th Summer School: Villa de Leyva, Colombia, July 6-23, 2009}},
  pp.~188--240, 2013, \href{https://doi.org/10.1017/CBO9781139208642.006}{DOI}.

\bibitem{Dixon:2012yy}
L.J.~Dixon, C.~Duhr and J.~Pennington, \emph{{Single-valued harmonic
  polylogarithms and the multi-Regge limit}}, {\emph{JHEP} {\bfseries 1210}
  (2012) 074} [\href{https://arxiv.org/abs/1207.0186}{{\ttfamily 1207.0186}}].

\end{thebibliography}\endgroup

\end{document}